\documentclass[aps,prc,twocolumn,10pt,superscriptaddress]{revtex4-2}
\usepackage{amsmath,amsfonts,amssymb}
\usepackage{graphicx}
\usepackage[dvipsnames]{xcolor}
\usepackage[normalem]{ulem}
\usepackage{xcolor}
\usepackage{hyperref}

\begin{document}

\title{Combined constraints on dark photons from high-energy collisions, cosmology, and astrophysics}

\author{A. W. Romero Jorge}
\email{jorge@itp.uni-frankfurt.de}
\affiliation{Frankfurt Institute for Advanced Studies (FIAS), Ruth Moufang Str. 1, 60438 Frankfurt, Germany}
 \affiliation{Institut f\"ur Theoretische Physik, Johann Wolfgang Goethe	University, Max-von-Laue-Str. 1, 60438 Frankfurt, Germany}
 \affiliation{Helmholtz Research Academy Hessen for FAIR (HFHF), GSI Helmholtz	Center for Heavy Ion Physics, Campus Frankfurt, 60438 Frankfurt, Germany.}

\author{L. Sagunski}
\affiliation{Institut f\"ur Theoretische Physik, Johann Wolfgang Goethe	University, Max-von-Laue-Str. 1, 60438 Frankfurt, Germany}

\author{Guan-Wen Yuan} 
\affiliation{Department of Physics, University of Trento, Via Sommarive 14, 38123 Povo (TN), Italy}
\affiliation{Trento Institute for Fundamental Physics and Applications (TIFPA)-INFN, Via Sommarive 14, 38123 Povo (TN), Italy}

\author{T. Song} 
\affiliation{GSI Helmholtzzentrum f\"ur Schwerionenforschung GmbH, Planckstraße 1, 64291 Darmstadt, Germany}

\author{E. Bratkovskaya}%
\affiliation{GSI Helmholtzzentrum f\"ur Schwerionenforschung GmbH, Planckstraße 1, 64291 Darmstadt, Germany}
 
\affiliation{Institut f\"ur Theoretische Physik, Johann Wolfgang Goethe	University, Max-von-Laue-Str. 1, 60438 Frankfurt, Germany}
\affiliation{Helmholtz Research Academy Hessen for FAIR (HFHF), GSI Helmholtz	Center for Heavy Ion Physics, Campus Frankfurt, 60438 Frankfurt, Germany.}

\date{\today}

\begin{abstract}
We investigate a dark sector coupled to the Standard Model (SM) through a
kinetically mixed dark photon $U$ associated with a new $U(1)'$ gauge symmetry.
Kinetic mixing, parametrized by $\varepsilon$, induces an
effective coupling to the electromagnetic current, while the dark photon
interacts with a stable dark matter (DM) particle $\chi$ through a dark gauge
coupling $g_\chi$,  defining a four-dimensional parameter space
$(m_U,\varepsilon,m_\chi,g_\chi)$.
Our analysis is based on the parton--hadron--string dynamics (PHSD) transport
approach, extended to include dark photon production and decay into dileptons
($U\!\to e^+e^-$). In PHSD, dark photons are produced in high-energy
collisions through Dalitz decays of light mesons
($\pi^0,\eta,\eta',\omega$), Delta-resonances
($\Delta\!\to N U$), direct vector meson decays
($\rho,\omega,\phi\!\to U$), kaon decays ($K^+\!\to\pi^+U$), and
$q\bar q\!\to U$ annihilation.
Building on previous PHSD benchmarks against dilepton data, we extract upper
limits on $\varepsilon^2(m_U,m_\chi,\alpha_\chi)$ in both the visible regime
($m_U<2m_\chi$), where $U\!\to e^+e^-$ dominates, and the invisible regime
($m_U>2m_\chi$), where $U\!\to\chi\bar\chi$ is kinematically open and suppresses
the dilepton branching fraction.
Cosmological and astrophysical constraints are incorporated in two
complementary ways. First, we compute the velocity-dependent self-interaction
cross section $\sigma/m_\chi$ for Yukawa-mediated SIDM and confront it with
bounds from dwarf galaxies, galaxy groups, and clusters. Second, we determine
thermal relic target curves by computing the relic abundance and requiring
$\Omega_{\rm DM}h^2\simeq 0.12$, consistent with \textit{Planck} measurements of
the cosmic microwave background.
Combining PHSD limits on $\varepsilon^2$ with relic density and self-interaction
requirements, we exclude regions of the $(m_\chi,m_U)$ plane for each DM
realization (Dirac fermion, Majorana fermion, or complex scalar) and identify
benchmark scenarios in which heavy-ion, cosmological, and astrophysical
constraints are simultaneously satisfied.
\end{abstract}

\maketitle

\section{Introduction}\label{sec:intro}

Dark matter (DM) is an essential ingredient of the standard cosmological model:
fits to the cosmic microwave background (CMB), large-scale structure, and other
cosmological probes imply that roughly one quarter of the total energy density
of the Universe resides in a non-luminous, non-baryonic component that
interacts predominantly via gravity \cite{Planck:2018vyg}.  
Its presence is required across a wide range of length scales.  On galactic
scales, stellar kinematics and neutral-hydrogen rotation curves show that the
circular velocity of spiral galaxies remains approximately flat far beyond the
visible stellar disc, pointing to extended DM halos
\cite{Feng:2010gw,deSalas:2019pee}.  
In galaxy clusters, strong and weak gravitational lensing in merging systems
such as the Bullet Cluster and Abell~520 reveals that most of the mass is
collisionless and spatially separated from the X-ray emitting intracluster gas
\cite{Clowe:2006eq,Mahdavi:2007yp}.  
Taken together, these observations robustly motivate new degrees of freedom
beyond the Standard Model (SM).

At the same time, high-resolution $N$-body simulations of cold,
collisionless DM exhibit tensions with observations on sub-galactic
scales.  
Cuspy inner density profiles and an overabundance of massive subhalos
predicted in $\Lambda$CDM do not always match the cored density
profiles and satellite populations of dwarf and low-surface-brightness galaxies \cite{Bullock:2017xww}.  
Self-interacting dark matter (SIDM), in which DM particles undergo elastic scattering with cross section per unit mass
$\sigma/m_\chi\sim 0.1$–$10~\mathrm{cm^2\,g^{-1}}$, provides a simple and well-motivated extension of $\Lambda$CDM that can alleviate the core–cusp and too-big-to-fail problems without spoiling its large-scale successes~\cite{Tulin:2017ara,Kaplinghat:2013yxa}.  
In realistic models, this often arises from a light mediator generating a Yukawa potential, which naturally leads to a velocity-dependent cross section: large at the low velocities of dwarf galaxies, and suppressed in the high-velocity regime characteristic of galaxy groups and clusters. 

Cosmology provides a complementary and independent handle on DM properties.
The CMB temperature and polarization anisotropies measured by \textit{Planck}
tightly constrain the present-day DM relic density to
$\Omega_{\rm DM}h^2\simeq 0.12$ \cite{Planck:2018vyg,WMAP:2012nax}.  
In thermal scenarios, this value is reproduced if the total (co)annihilation
rate of DM in the early Universe is of order
$\langle\sigma v\rangle\sim 10^{-26}~\mathrm{cm^3\,s^{-1}}$ \cite{Kolb:1990vq}, thereby
linking cosmological observations to the microscopic couplings of the dark
sector.  Any viable model must therefore accommodate both the relic density
requirement and the astrophysical constraints on self-interactions.

Despite intense efforts, direct and indirect searches have not yet yielded a
conclusive detection of DM.  Underground detectors probing nuclear and
electronic recoils, $\gamma$-ray and cosmic-ray observations of the Galactic
halo, and collider searches at the LHC have excluded large regions of parameter
space, but still leave ample room for light mediators and non-minimal dark
sectors.  This motivates exploring frameworks in which the same mediator
controls early-Universe annihilation, potential self-interactions, and
laboratory signatures.

A minimal and well-motivated way to couple a dark sector to the Standard Model
(SM) is via renormalizable \emph{portals}
\cite{Alexander:2016aln,Battaglieri:2017aum,Agrawal:2021dbo,Yuan:2022nmu}.  
These are dimension-four operators that connect SM fields to new gauge-singlet
degrees of freedom.
In this work we focus on the vector portal, in which a new Abelian $U(1)'$
gauge boson kinetically mixes with hypercharge.  The interaction
$
\mathcal{L}\supset \frac{\varepsilon}{2}\,B_{\mu\nu}F^{\prime\mu\nu}
$
first proposed by Holdom~\cite{Holdom:1985ag} induces a coupling of the
associated gauge boson, often referred to as a dark photon, hidden photon, or
$U$ boson to the electromagnetic current with strength $\varepsilon e$.  
The dark photon can acquire a mass $m_U$ through a Stueckelberg term or a dark
Higgs mechanism, and its coupling to charged SM particles is then set by
$\varepsilon^{2}$, which controls its partial widths into $e^{+}e^{-}$,
$\mu^{+}\mu^{-}$, and hadronic states
\cite{Fayet:1980ad,Fayet:2004bw,Boehm:2003hm,Pospelov:2007mp,Batell:2009di,Batell:2009yf}.  
For $\varepsilon$ in the range $10^{-7}$-$10^{-2}$ such states can be long-lived
yet remain compatible with precision electroweak data, while still being
accessible to dedicated searches.

An important phenomenological advantage of the vector portal is that a dark
photon can be radiated from any electromagnetic current.  In the low-mass
region, this implies a rich set of hadronic production channels: Dalitz decays
of light pseudoscalars ($\pi^0$, $\eta$, $\eta'$), radiative decays of baryon
resonances such as $\Delta\to N\gamma^\ast$, and direct conversions of vector
mesons $\rho$, $\omega$, and $\phi$.  Searches typically look for a narrow
resonance in the dilepton invariant-mass spectrum above the smooth SM
background
\cite{Agakishiev:2013fwl,Beacham:2019nyx,Billard:2021uyg}.  
A broad experimental programme ranging from fixed-target 
and beam-dump
experiments to flavor factories and LHC detectors has excluded
$\varepsilon^{2}\gtrsim 10^{-6}$ for dark photon masses between a few tens of
MeV and several GeV
\cite{Fabbrichesi:2020wbt,APEX:2011dww,HPS:2018xkw,
Merkel:2014avp,Batley:2015lha,
BaBar:2009lbr,BaBar:2014zli,
KLOE-2:2014qxg,KLOE-2:2012lii,KLOE-2:2018kqf,
LHCb:2017trq,Aaij:2019bvg,CMS:2023hwl}.  
Nevertheless, important gaps remain, in particular near the $\rho$, $\omega$,
and $\phi$ resonances and in regions where laboratory, beam-dump, and
astrophysical constraints lose sensitivity. These gaps motivate new dedicated searches to extend sensitivity in the MeV-GeV range
\cite{Duarte:2022feb,Bechtle:2024atq}.

Heavy-ion collisions offer a complementary window on the vector portal.  They
create hot and dense strongly interacting matter, produce a large number of light mesons
and baryon resonances, and, at collider energies, form a thermalized
quark-gluon plasma (QGP).  Both the hadronic and partonic stages emit virtual
photons and, in the presence of kinetic mixing, dark photons, which subsequently
decay into dileptons.  In our previous studies, we implemented this mechanism
within the parton-hadron-string dynamics (PHSD) transport framework to derive
upper limits on $\varepsilon^{2}(m_U)$ in the \emph{visible} regime, where the
decay $U\to e^{+}e^{-}$ dominates and the dark channel is kinematically
inaccessible.  Those works focused on dark photons produced through hadronic and
partonic sources, demonstrating that high-precision dilepton spectra from
$p+p$, $p+A$, and $A+A$ collisions provide a competitive probe of dark photons
across the MeV-GeV range
\cite{Schmidt:2021hhs,Bratkovskaya:2022cch,RomeroJorge:2024eky,Jorge:2024ris}.

In the present work, we embed this dark photon phenomenology into a more general
dark sector framework and substantially extend the scope of the analysis.  We consider a dark photon coupled to a stable DM particle $\chi$ with dark gauge
coupling $g_{\chi}$, exploring both fermionic (Dirac and
Majorana) and complex scalar realizations.  The resulting
parameter space $(m_U,\varepsilon,m_\chi,g_{\chi})$ links heavy-ion observables
to cosmology and astrophysics through thermal freeze-out and self-interacting DM
dynamics.  
Throughout this work we distinguish between the mass range directly probed by
dilepton measurements in heavy-ion collisions and the broader parameter space
relevant for self-interacting dark matter.
The PHSD dilepton analysis constrains mediators with $m_U>2m_e$, for which the
two-body decay $U\to e^+e^-$ is kinematically open, and up to a few GeV. In the SIDM part we also explore
ultra-light mediators, including the sub-MeV domain, when computing Yukawa
self-interaction cross sections.
Our main advances compared to our previous PHSD study \cite{Jorge:2025gph} are:
\begin{itemize}
    \item[(i)] \textbf{Visible and invisible regimes:}
    we extend the PHSD-based extraction of $\varepsilon^{2}(m_U)$ to the
    invisible regime where $U\to\chi\bar\chi$ is open, deriving upper
    limits on the kinetic mixing.

    \item[(ii)] \textbf{Astrophysical self-interaction constraints:}
    we compute the velocity-dependent self-interaction cross section
    $\sigma/m_\chi$ using \textsc{CLASSICS} and confront it with SIDM
    bounds from dwarf galaxies, galaxy groups, and clusters, thereby excluding
    portions of the $(m_\chi,m_U)$ plane where self-interactions are either too
    weak to alleviate small-scale tensions or too strong to be compatible with
    high-velocity systems.

    \item[(iii)] \textbf{Thermal relic targets and invisible limits:}
    employing \textsc{ReD-DeLiVeR}, we determine thermal relic target curves in
    $(m_\chi,m_U,\varepsilon,g_{\chi})$ space for Dirac, Majorana, and complex
    scalar DM, and combine them with our PHSD limits on the invisible kinetic
    mixing to rule out parts of the parameter space where a standard thermal
    history would require a mixing already excluded by heavy-ion dilepton data.

    \item[(iv)] \textbf{Benchmark scenarios:}
    we identify three representative benchmark points in the $(m_\chi,m_U)$
    plane that simultaneously satisfies heavy-ion, cosmological, and astrophysical
    constraints, and are thus particularly promising targets for future
    experimental searches.
\end{itemize}

The paper is structured as follows.  In Sec.~\ref{sec:DPmodel}, we introduce the
dark photon framework with a generic dark sector, specifying the kinetic mixing
Lagrangian, the dark currents, and mass terms for Dirac, Majorana, and complex
scalar dark matter, and the resulting visible and invisible partial widths and
branching ratios.  In Sec.~\ref{sec:phsd}, we summarize the PHSD transport
approach and its implementation of standard model dilepton sources, and we
describe how dark photons are produced in hadronic and partonic channels, and
how visible and invisible limits on $\varepsilon^2(m_U)$ are extracted from
dilepton spectra.  Section~\ref{sec:sidm} is devoted to self-interacting dark
matter: we compute Yukawa-mediated self-interaction cross sections with
\textsc{CLASSICS}, construct the effective cross section $\sigma_{\rm eff}/m_\chi$,
and compare to astrophysical constraints from dwarfs, galaxies, groups, and
clusters.  In Secs.~\ref{subsec:thermal_relic} and \ref{subsec:reddeliver} we
discuss the thermal relic abundance, use \textsc{ReD-DeLiVeR} to obtain
thermal relic target curves for the different DM spins, and derive their
scaling in the nonrelativistic limit.  In Sec.~\ref{sec:mu_mx_planes} we
combine heavy-ion, cosmological, and astrophysical constraints in the
$(m_\chi,m_U)$ plane, identify regions where thermal targets are excluded by
invisible PHSD limits, and define three representative benchmark scenarios.
Finally, Sec.~\ref{sec:summary} summarizes our main results and outlines
prospects for future experimental tests. 
We work in natural units $\hbar=c=k_B=1$ unless stated otherwise. Masses are quoted in MeV/GeV, velocities in km\,s$^{-1}$, and self-interaction cross sections per unit mass in cm$^2$\,g$^{-1}$.

\section{Modeling of dark photon production with a generic dark sector}
\label{sec:DPmodel}

A minimal and well-motivated way to couple a hidden sector to the Standard Model
(SM) is through renormalizable \emph{portals}, i.e.\ dimension-four operators
linking SM fields to new gauge-singlet degrees of freedom.
Prominent examples include the Higgs portal (spin-0 scalar mediator),
the neutrino portal (spin-$1/2$ singlet fermion), the axion/pseudoscalar portal
(spin-0 pseudoscalar), and the \emph{vector portal}, in which a new Abelian
spin-1 gauge boson kinetically mixes with hypercharge
\cite{Holdom:1985ag,Batell:2009di,Essig:2013lka,Alexander:2016aln}.
In this work we focus on the vector portal.  At energies well below the $Z$ mass,
kinetic mixing induces an effective coupling of the dark photon $U$ 
to the electromagnetic current with strength $\varepsilon e$.

To allow for invisible decays of the mediator, we consider a generic dark sector
containing a particle charged under a dark $U(1)'$ gauge symmetry.
We assume an elastic setup with a single stable
dark sector state $\chi$ (or $\varphi$ for a complex scalar field), so that invisible decays proceed dominantly
via $U\to\chi\bar\chi$ (or $U\to\varphi\varphi^\dagger$) when kinematically allowed.
The interaction strength is parametrized by $g_\chi$ (or equivalently
$\alpha_\chi \equiv g_\chi^2/4\pi$), and the minimal renormalizable Lagrangian is
\begin{eqnarray}
\mathcal{L} &=& \mathcal{L}_{\rm SM}
-\frac{1}{4}F'_{\mu\nu}F'^{\mu\nu}
+\frac{\varepsilon}{2}\,B_{\mu\nu}F'^{\mu\nu}
+\frac{1}{2}m_U^2\,A'_\mu A'^{\mu}
\nonumber\\
&&\quad
- g_\chi\, A_\mu^\prime\, J^{\mu}_\chi
+ \mathcal{L}_{\rm mass}^{\chi},
\label{eq:lagrangian_dp_dm}
\end{eqnarray}
where $B_{\mu\nu}$ and $F'_{\mu\nu}$ denote the hypercharge and dark
field strength tensors, respectively. The dark photon field is $A'_\mu$ and its
mass is $m_U$. Throughout, we parametrize the dark sector interaction strength by
$\alpha_\chi \equiv g_\chi^2/(4\pi)$ and use the same $g_\chi$ normalization for
all spin assignments; the different numerical prefactors in the partial widths
below arise solely from the Lorentz structure of the current and
spin/identical-particle statistics.

After electroweak symmetry breaking and diagonalization of the gauge-kinetic
terms, the dominant low-energy interaction is
$\varepsilon e\,A'_\mu J_{\rm EM}^\mu$, up to
$\mathcal{O}(m_U^2/m_Z^2)$ corrections from the induced $Z$-current coupling \cite{Holdom:1985ag,Essig:2013lka}.

The structure of the dark current $J^\mu_\chi$ depends on the spin of the dark
sector state and is given by \cite{Krnjaic:2025noj}
\begin{equation}
J_\chi^\mu =
\left\{
\begin{array}{ll}
\bar{\chi}\gamma^\mu \chi,
& \text{Dirac fermion}, \\[6pt]
\frac{1}{2}\bar{\chi}\gamma^\mu\gamma^5 \chi,
& \text{Majorana fermion}, \\[6pt]
i\!\left(\varphi^\dagger \partial^\mu \varphi
 - (\partial^\mu \varphi^\dagger)\varphi\right),
& \text{complex scalar},
\end{array}
\right.
\end{equation}
and the corresponding mass terms read
\begin{equation}
\mathcal{L}_{\rm mass}^\chi =
\left\{
\begin{array}{ll}
-\, m_\chi\, \bar{\chi}\chi,
& \text{Dirac fermion}, \\[6pt]
-\, \dfrac{1}{2}\, m_\chi\, \bar{\chi}\chi,
& \text{Majorana fermion}, \\[6pt]
-\, m_\chi^2 \, \varphi^\dagger \varphi,
& \text{complex scalar}.
\end{array}
\right.
\end{equation}

Production of $U$ from SM sources in hadronic and nuclear environments proceeds through kinetic mixing and therefore is universal at leading order, differing from the corresponding virtual-photon rate  by an overall factor of $\varepsilon^2$. Decays are controlled by the visible width into SM states and the invisible width into dark states. For charged leptons,
\begin{equation}
\Gamma(U\!\to\!\ell^+\ell^-)=\frac{1}{3}\,\alpha\,\varepsilon^2\,m_{U}\!
\left(1+\frac{2m_\ell^2}{m_{U}^2}\right)\!
\sqrt{1-\frac{4m_\ell^2}{m_{U}^2}},
\label{eq:width_ll_gen}
\end{equation}
and the hadronic width is incorporated via the data-driven $R$-ratio (see Ref. \cite{Schmidt:2021hhs,Agakishiev:2013fwl}),
\begin{eqnarray}
\Gamma_{\rm had}(m_{U}) &=& \Gamma_{\mu\mu}(m_{U})\;R(m_{U}),
\\
R(s)&\equiv&
\frac{\sigma(e^+e^-\!\to\!\text{hadrons})}{\sigma(e^+e^-\!\to\!\mu^+\mu^-)}
\bigg|_{s=m_{U}^2}.
\label{eq:width_had_gen}
\end{eqnarray}
The total visible width is obtained by summing all kinematically accessible SM
final states,
\begin{equation}
\Gamma_{\rm vis}(m_{U})
= \sum_{\ell=e,\mu,\tau}\Gamma(U\!\to\!\ell^+\ell^-)
\;+\;\Gamma_{\rm had},
\label{eq:vis_width_sum}
\end{equation}
and the branching ratio into electrons becomes
\begin{equation}
{\rm Br}(U\!\to e^+e^-) \;=\;
\begin{cases}
\dfrac{\Gamma(U\!\to e^+e^-)}{\Gamma_{\rm vis}},
& m_U < 2m_{\chi}, \\[10pt]
\dfrac{\Gamma(U\!\to e^+e^-)}{\Gamma_{\rm vis}+\Gamma_{\rm inv}},
& m_U \ge 2m_{\chi},
\end{cases}
\label{BrVisINV}
\end{equation}
where $\Gamma_{\rm inv}=0$ below threshold.

The invisible width $\Gamma_{\rm inv}$ depends on the Lorentz structure of the
dark current and on the spectrum of dark states.  We consider three benchmark
dark matter realizations:

\medskip
\noindent
(i) \textbf{Dirac fermion} $\chi$ with vector coupling
$J_\chi^\mu=\bar\chi\gamma^\mu\chi$, for which
\begin{equation}
\Gamma(U\!\to\!\bar\chi\chi)=
\frac{1}{3}\,\alpha_\chi \,m_{U}\!
\left(1+\frac{2m_\chi^2}{m_{U}^2}\right)
\sqrt{1-\frac{4m_\chi^2}{m_{U}^2}}.
\label{eq:width_dirac}
\end{equation}

\medskip
\noindent
(ii) \textbf{Majorana fermion} $\chi$ with axial-vector current
$J_\chi^\mu=\bar\chi\gamma^\mu\gamma^5\chi$.
Since the vector current vanishes identically for a Majorana field, the
invisible width is
\begin{equation}
\Gamma(U\!\to\!\chi\chi)=
\frac{1}{6}\,\alpha_\chi\,m_{U}\!
\left(1-\frac{4m_\chi^2}{m_{U}^2}\right)^{3/2},
\label{eq:width_majorana}
\end{equation}
\medskip
\noindent
(iii) \textbf{Complex scalar} $\varphi$ with current
$J_\varphi^\mu = i\varphi^\dagger\!\overleftrightarrow{\partial^\mu}\!\varphi$, yielding \footnote{For notational simplicity we denote the dark matter mass by $m_\chi$ throughout,
also in the complex-scalar case $\varphi$, i.e.\ $m_\varphi\equiv m_\chi$.
}
\begin{equation}
\Gamma(U\!\to\!\varphi \varphi^\dagger)=
\frac{1}{12}\,\alpha_\chi\,m_{U}\!
\left(1-\frac{4m_\chi^{\,2}}{m_{U}^{\,2}}\right)^{3/2}.
\label{eq:width_scalar}
\end{equation}

For Majorana dark matter, the vector current vanishes identically, so the leading
renormalizable interaction with a spin-1 mediator is axial. This yields the
near-threshold scaling $\Gamma_{\rm inv}\propto(1-4m_\chi^2/m_U^2)^{3/2}$ and the
corresponding normalization in Eq.~(\ref{eq:width_majorana}).

\subsection{Visible and Invisible Decays of the Dark Photon}
\label{subsec:vis_inv_decays}

\begin{figure*}[t]
    \centering
\includegraphics[width=0.99\linewidth]{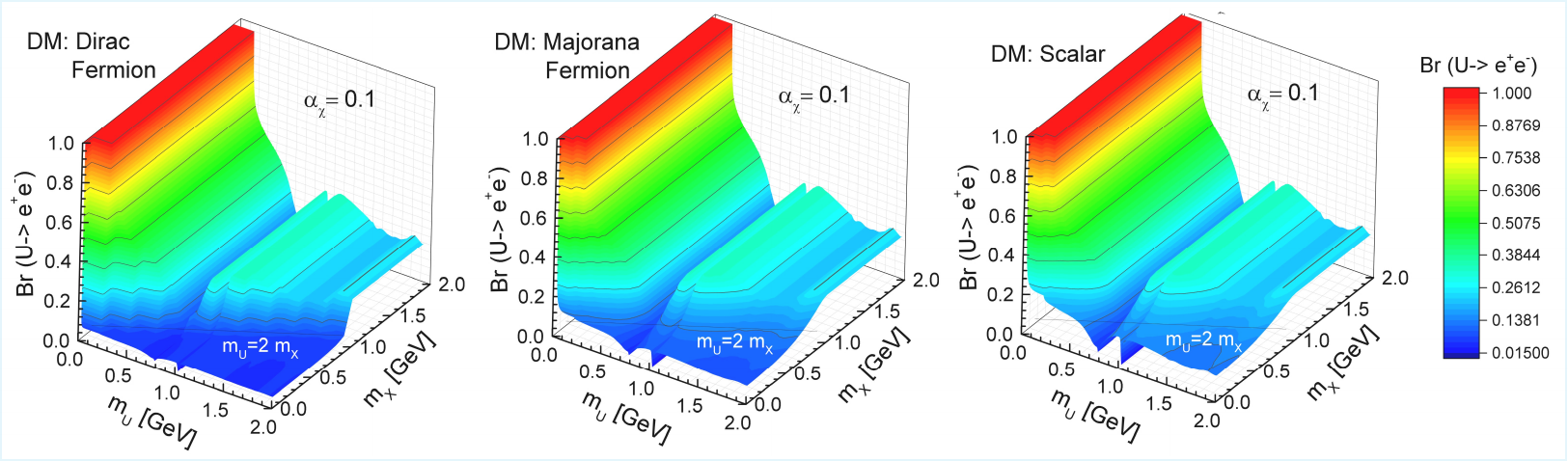}
    \caption{
Branching fraction $\mathrm{Br}(U\to e^{+}e^{-})$ as a function of the mediator
mass $m_{U}$ and the dark matter mass $m_{\chi}$ for $\alpha_{\chi}=0.1$, shown
for three benchmark dark matter candidates: a Dirac fermion (left), a Majorana
fermion (center), and a complex scalar (right).
A sharp drop occurs along the kinematic threshold $m_{U}=2m_{\chi}$, where the
invisible decay $U\to\chi\bar{\chi}$ (or $U\to \varphi\varphi^\dagger$) opens and rapidly
dominates the total width.
Below threshold ($m_{U}<2m_{\chi}$), decays proceed exclusively into
standard model final states and the dileptonic branching fraction remains close
to unity, modulated by hadronic contributions near the $\rho$, $\omega$, and
$\phi$ resonances.
Differences among the three panels reflect the distinct spin structures and
threshold behaviour of the invisible partial width $\Gamma_{\rm inv}$ for
fermionic and scalar dark matter. The visible widths are taken from Ref.~\cite{Ilten:2015hya}, while $\Gamma_{\rm inv}$ is computed from
Eqs.~(\ref{eq:width_dirac})-(\ref{eq:width_scalar}).
}
\label{BR_U_ee_3d}
\end{figure*}

To illustrate how visible and invisible decay channels compete across the
dark sector parameter space, Fig.~\ref{BR_U_ee_3d} displays the dileptonic
branching fraction $\mathrm{Br}(U\!\to e^{+}e^{-})$ in the
$(m_{U},m_{\chi})$ plane for a fixed coupling $\alpha_{\chi}=0.1$.  The three
panels correspond to the benchmark spin assignments considered throughout this
work: Dirac fermion, Majorana fermion, and complex scalar dark matter.

The visible branching fraction is determined by the competition between the
visible width $\Gamma_{\rm vis}(m_{U})$ and the invisible width
$\Gamma_{\rm inv}(m_{U},m_{\chi},\alpha_{\chi})$,
so that the transition between the \emph{visible} and \emph{invisible} regimes
is controlled primarily by the kinematic opening of the dark channel at
$m_{U}=2m_{\chi}$.

\paragraph*{Invisible regime ($m_{U}>2m_{\chi}$).}
In each panel, a pronounced transition appears along the kinematic boundary
$m_{U}=2m_{\chi}$, where the invisible decay $U\!\to\chi\bar{\chi}$ (or
$U\!\to XX^{\dagger}$ for scalar DM) becomes allowed and typically dominates the
total width.  Parametrically, close to threshold one has
\begin{equation}
  \Gamma_{\rm inv} \;\propto\; \alpha_{\chi}\,m_{U}\,
  \Big(1-\frac{4m_{\chi}^{2}}{m_{U}^{2}}\Big)^{p},
  \label{eq:GammaInv_scaling}
\end{equation}
with $p=\tfrac{1}{2}$ for a Dirac fermion (unsuppressed $s$-wave phase space) and
$p=\tfrac{3}{2}$ for Majorana and complex-scalar final states (threshold
suppression from the corresponding coupling/phase-space structure).  As a
result, once $m_{U}>2m_{\chi}$ the visible branching fraction can drop by orders
of magnitude, which is reflected in the steep decrease of
$\mathrm{Br}(U\!\to e^{+}e^{-})$ to $\mathcal{O}(10^{-2})$ or below over large
regions of parameter space.

\paragraph*{Visible regime ($m_{U}<2m_{\chi}$).}
For $m_{U}<2m_{\chi}$ the dark channel is closed ($\Gamma_{\rm inv}=0$) and the
dark photon decays exclusively into standard model states.  In this region
$\mathrm{Br}(U\!\to e^{+}e^{-})$ remains close to unity over most of the plane,
with narrow modulations from QCD resonances.  These features originate from the
hadronic contribution to the total width,
$\Gamma_{\rm had}(m_{U})\propto R(m_{U})\,\Gamma_{\mu\mu}(m_{U})$, which is
enhanced near the $\rho$, $\omega$, and $\phi$ poles and temporarily reduces
the relative weight of the dilepton mode even in the absence of invisible
decays.

The differences between the three spin scenarios are therefore driven mainly by
the threshold behaviour of $\Gamma_{\rm inv}$: Dirac dark matter, with an
unsuppressed vector coupling, typically exhibits the most abrupt loss of
visibility above threshold, while the Majorana and scalar cases display a
somewhat smoother onset due to the stronger threshold suppression in
Eq.~(\ref{eq:GammaInv_scaling}).  Overall, Fig.~\ref{BR_U_ee_3d} identifies the
regions in the $(m_{U},m_{\chi})$ plane where dilepton measurements are
intrinsically most sensitive (visible regime) and those where invisible decays
render the mediator effectively hidden, motivating complementary cosmological
and astrophysical constraints.

\begin{figure*}[t]
    \centering
    \includegraphics[width=0.49\linewidth]{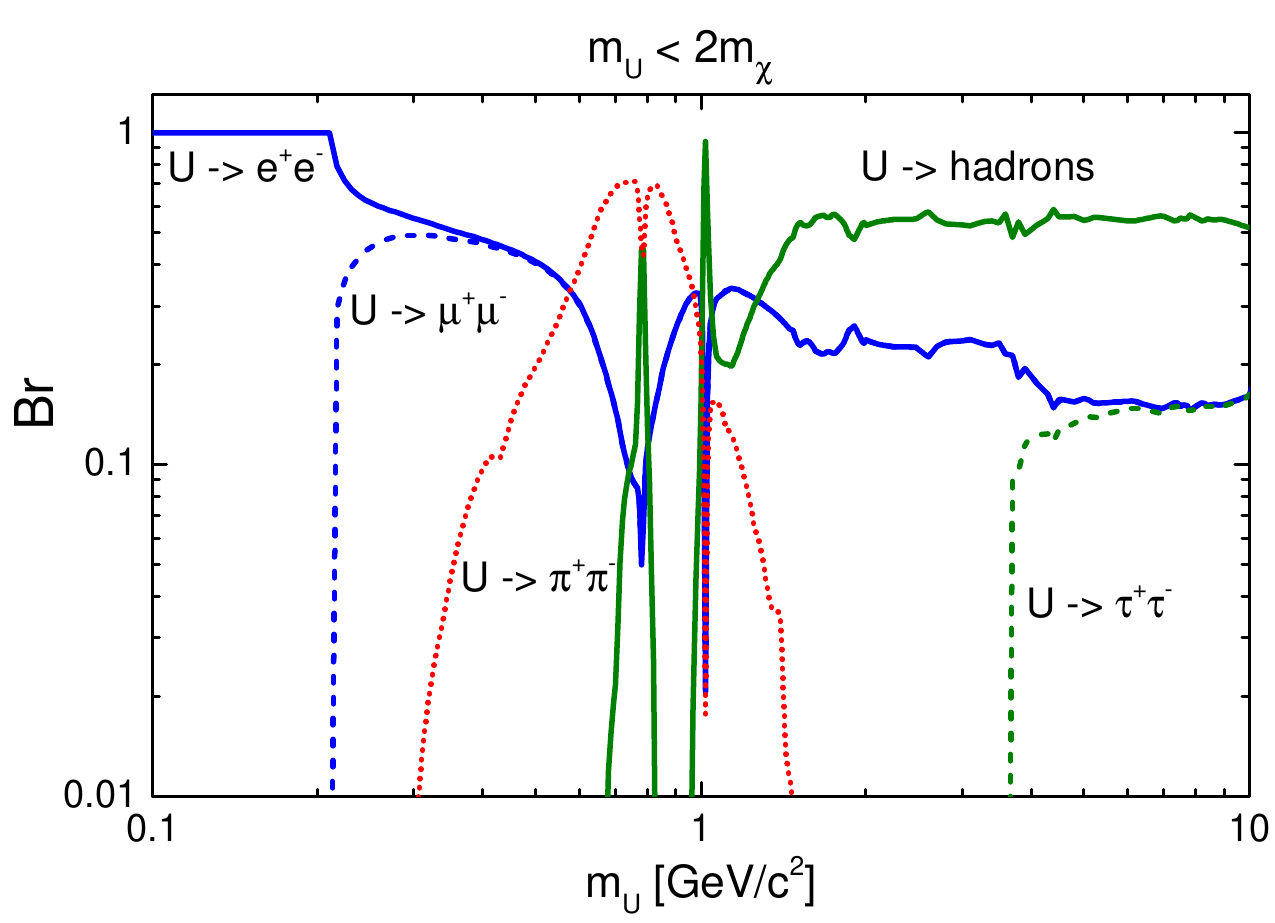}
    \includegraphics[width=0.49\linewidth]{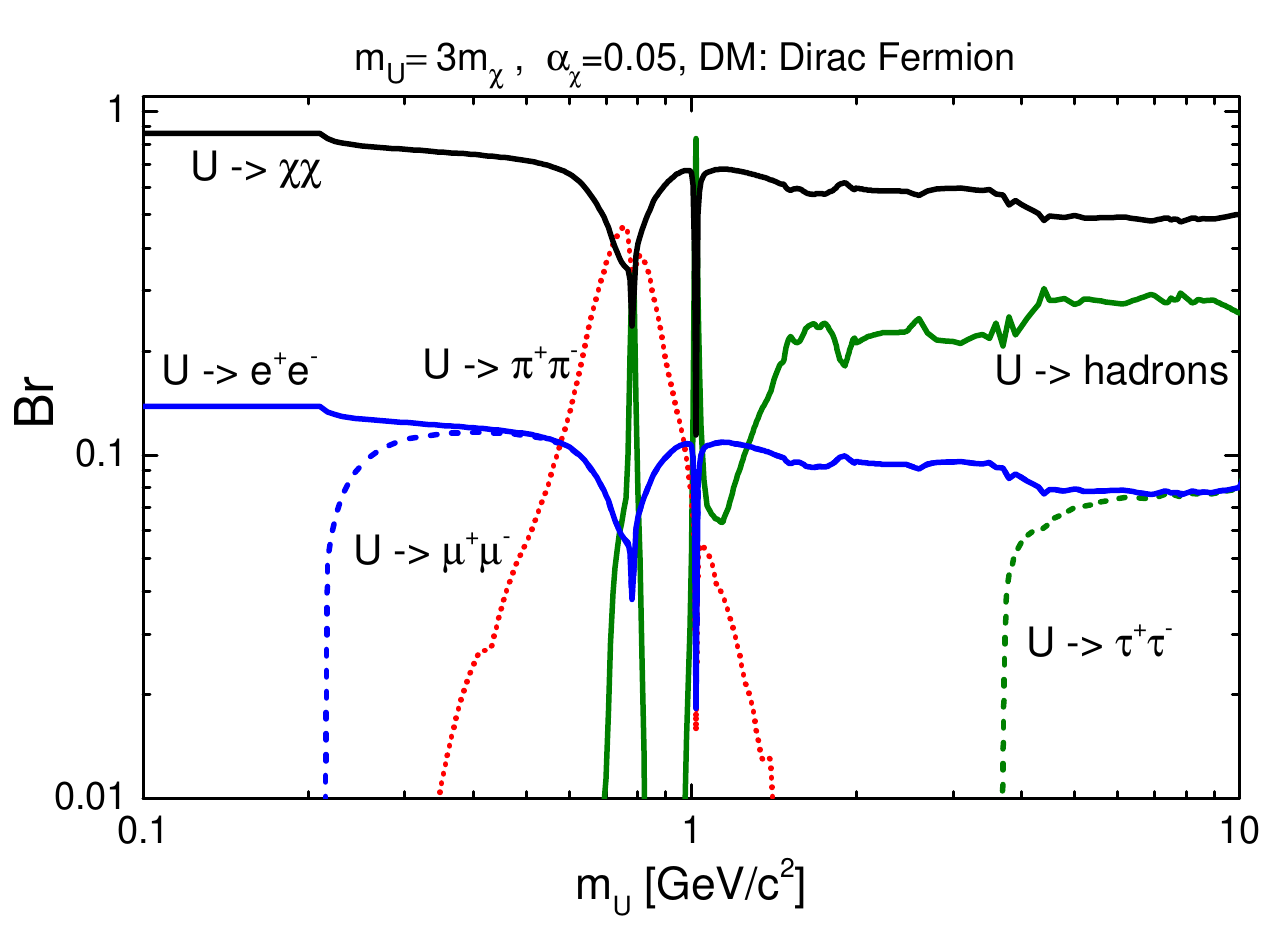}
    \caption{
Branching fractions of the dark photon $U$ as a function of its mass $m_{U}$ in
the visible and invisible regimes.
Left: \emph{visible regime}, $m_{U}<2m_{\chi}$, where decays into the dark sector
are kinematically forbidden and the total width is shared among standard model
final states.  We show $U\!\to e^{+}e^{-}$ (solid blue),
$U\!\to\mu^{+}\mu^{-}$ (dashed blue), $U\!\to\tau^{+}\tau^{-}$ (dotted green),
the inclusive hadronic mode (solid green), and the exclusive $\pi^{+}\pi^{-}$
channel (dotted red).
Right: \emph{invisible regime} for fermionic dark matter, illustrated for
$R\equiv m_{U}/m_{\chi}=3$ and $\alpha_{\chi}=0.05$.  In this benchmark the
invisible decay $U\!\to\chi\bar{\chi}$ (solid black) dominates over most of the
mass range, while the standard model channels retain the same threshold and
resonance structure as in the visible case but are globally suppressed.
Prominent QCD resonances ($\rho$, $\omega$, $\phi$, \dots) produce localized
features in the hadronic curves and induce corresponding dips and peaks in the
leptonic branching fractions. The visible (hadronic and leptonic) widths entering $\Gamma_{\rm vis}(m_U)$ are
taken from Ref.~\cite{Ilten:2015hya}, while $\Gamma_{\rm inv}$ is computed from
Eqs.~(\ref{eq:width_dirac})-(\ref{eq:width_scalar}).}
\label{BR_U_ee_2d} 
\end{figure*}

Figure~\ref{BR_U_ee_2d} provides a two-dimensional view of the decay pattern
across the GeV mass range.  The left panel corresponds to the visible regime,
$m_{U}<2m_{\chi}$, where phase space enforces the expected hierarchy of leptonic
modes: $U\!\to e^{+}e^{-}$ dominates just above threshold, followed by
$\mu^{+}\mu^{-}$ once $m_{U}>2m_{\mu}$, and $\tau^{+}\tau^{-}$ for
$m_{U}>2m_{\tau}$.  The hadronic contribution turns on near the two-pion
threshold and exhibits pronounced resonance structure near the $\rho$, $\omega$,
and $\phi$ poles, which propagates directly into the leptonic branching
fractions.

The right panel shows the invisible regime for a representative fermionic
benchmark with $R=3$ and $\alpha_{\chi}=0.05$.  Here $U\!\to\chi\bar{\chi}$ is
open and typically dominates, so the invisible branching fraction stays close
to unity except in narrow windows where hadronic resonances transiently enhance
$\Gamma_{\rm vis}$.  Consequently, the standard model channels exhibit the same
threshold and resonance pattern as in the visible case, but are globally
suppressed by the large invisible width.

Having established how the dark photon decays into visible and invisible final
states, we now turn to its production in the hot and dense environment created
in relativistic nuclear collisions.

\section{Standard matter and dark photon production in PHSD}
\label{sec:phsd}

\subsection{PHSD approach}

The parton–hadron–string dynamics (PHSD) model is a microscopic, non–equilibrium
transport approach that follows both hadronic and partonic degrees of freedom
throughout the full space–time evolution of a relativistic heavy-ion
collision~\cite{Cassing:2008sv,Cassing:2008nn,Cassing:2009vt,Bratkovskaya:2011wp,Linnyk:2015rco,Moreau:2019vhw}.  
Starting from the first off-shell $NN$ encounters, PHSD describes the string
excitation and fragmentation, the formation and expansion of the quark–gluon
plasma (QGP), the subsequent hadronization near the QCD cross-over, and the
late hadronic rescattering and decays within a single unified framework.

The dynamics are obtained from the Cassing–Juchem off-shell transport equations,
which arise from a first-order gradient expansion of the Kadanoff–Baym
equations in test-particle representation and explicitly retain finite spectral
functions~\cite{Cassing:1999wx,Juchem:2004cs}.  
At high energy density, color strings melt into off-shell quarks, antiquarks
and gluons described by the Dynamical Quasi-Particle Model (DQPM), whose
complex self-energies and effective couplings are tuned to reproduce the
lattice-QCD equation of state and transport coefficients over the
$(T,\mu_B)$ plane~\cite{Cassing:2007yg,Moreau:2019vhw,Soloveva:2019xph}.  
As the system cools and dilutes, parton spectral functions sharpen and
hadronization proceeds continuously; the ensuing hadronic system is then
propagated with off-shell HSD dynamics.  
PHSD has been benchmarked from SIS to RHIC and LHC energies for a wide range of
hadronic, photon, and dilepton observables (see, e.g.,
Refs.~\cite{Linnyk:2015rco,Moreau:2021clr} and the detailed discussion in
Ref.~\cite{Jorge:2025gph}).

\subsection{Dilepton production from standard model sources}
\label{sec:dileptonSM}

Throughout the full evolution of the collision, PHSD tracks the emission of
virtual photons, which subsequently convert into dilepton pairs.  
All established standard model sources are included as in
Refs.~\cite{Jorge:2025gph,Linnyk:2015rco,Rapp:2013nxa}:

\begin{itemize}
    \item \emph{Low-mass hadronic channels ($M_{ee}\!\lesssim\!1$~GeV/$c^2$):}  
    Dalitz decays of pseudoscalar mesons 
    $\pi^{0},\eta,\eta'\!\to\gamma e^{+}e^{-}$ and baryon resonances 
    $\Delta\!\to N e^{+}e^{-}$, $N^\ast\!\to N e^{+}e^{-}$,  
    Dalitz transitions of vector and axial–vector mesons 
    $\omega\!\to\pi^{0} e^{+}e^{-}$, $a_{1}\!\to\pi e^{+}e^{-}$,  
    direct two–body decays of light vector mesons  
    $\rho,\omega,\phi\!\to e^{+}e^{-}$,  
    as well as nucleon–nucleon and meson–nucleon bremsstrahlung  
    $NN\!\to NN e^{+}e^{-}$, $\pi N\!\to\pi N e^{+}e^{-}$,  
    all implemented with off–shell spectral functions and medium–modified widths  
    as in Refs.~\cite{Bratkovskaya:2007jk,Bratkovskaya:2013vx,Linnyk:2015rco,Jorge:2025gph}.
    
    \item \emph{Partonic emission from the QGP:}  
    annihilation and Compton–like processes  
    $q\bar q\!\to\gamma^\ast$, 
    $q\bar q\!\to g\gamma^\ast$, 
    $qg(\bar q g)\!\to q(\bar q)\gamma^\ast$,  
    evaluated with DQPM propagators and a temperature– and density–dependent
    running coupling, thereby incorporating finite parton masses and widths, and
    reproducing the lattice-QCD equation of state~\cite{Moreau:2019vhw,Moreau:2021clr,Jorge:2025gph}.
    
    \item \emph{Intermediate- and high-mass region:}  
    correlated semi\-leptonic decays of open-charm and open-beauty mesons 
    (from $D\bar D$, $B\bar B$ production) dominate for 
    $1\!\lesssim\!M_{ee}\!\lesssim\!3$~GeV/$c^2$, 
    with heavy-quark transport and energy loss treated as in 
    Ref.~\cite{Song:2018xca}.  
    At still higher masses, Drell–Yan–like $q\bar q\!\to\gamma^\ast$ processes in
    initial hard scatterings are included following the standard PHSD
    implementation~\cite{Linnyk:2015rco,Jorge:2025gph}.
\end{itemize}

Radiation from hadronic resonances is implemented via the continuous
“shining’’ procedure: each off-shell resonance contributes to the dilepton
yield along its trajectory according to its instantaneous electromagnetic width
and local medium properties.  
This method consistently accounts for finite lifetimes, off-shell masses, and
in-medium modifications of all emitters; a detailed description and validation
are given in
Refs.~\cite{Bratkovskaya:2007jk,Bratkovskaya:2013vx,Galatyuk:2015pkq,Jorge:2025gph}.

\subsection{Dark photon production in PHSD}
\label{sec:DPproduction}

Dark photon production is implemented as an extension of PHSD by introducing an on-shell dark photon $U$ coupled through kinetic mixing and assuming a narrow width, such that each $U$ contributes to a single invariant-mass bin in the dilepton spectrum~\cite{Schmidt:2021hhs,Bratkovskaya:2022cch,Jorge:2025gph}.  

We include the following production channels in PHSD:
\begin{align}
\pi^{0},\eta,\eta^{\prime} &\to \gamma\,U, \label{decay1}\\
\omega &\to \pi^{0}\,U, \\
\Delta &\to N\,U, \\
\rho,\omega,\phi &\to U, \\
K^{+} &\to \pi^{+}\,U, \label{decay9}\\
q\bar q &\to U. \label{decay10}
\end{align}
For Dalitz-type channels $m\to\gamma U$ with $m=\pi^0,\eta,\eta'$ and for
$\omega\to\pi^0 U$, we obtain the rates from the corresponding virtual-photon
decays via the partial-width ratios
$\Gamma_{m\to\gamma U}/\Gamma_{m\to\gamma\gamma}$ and
$\Gamma_{\omega\to\pi^0 U}/\Gamma_{\omega\to\pi^0\gamma}$, which directly yield
the $m\to XU$ branching ratios as in
Refs.~\cite{Batell:2009yf,Batell:2009di,Gorbunov:2024nph}.
The $\Delta\to N U$ contribution is evaluated with the $\Delta$ spectral
function and a mass-dependent total width following
Refs.~\cite{Agakishiev:2013fwl,Bratkovskaya:2013vx,Jorge:2025gph}.
Direct decays of vector mesons $V=\rho,\omega,\phi$ are implemented by replacing
the virtual photon by a dark photon with effective coupling
$\alpha'=\alpha\,\varepsilon^2$ \cite{Batell:2009di}.
For the kaon mode $K^+\to\pi^+U$ we adopt the branching ratio from
Ref.~\cite{Pospelov:2008zw}.
The partonic channel $q\bar q\to U$ is obtained by rescaling the PHSD
$q\bar q\to\gamma^\ast\to e^+e^-$ yield with $\varepsilon^2$, exploiting the
off-shell nature of DQPM partons which kinematically allows single-$U$
production \cite{Berlin:2018pwi,Jorge:2025gph}.

Production from SM sources in hadronic and nuclear environments proceeds
entirely through kinetic mixing and is therefore universal up to the overall
factor $\varepsilon^2$ relative to the corresponding virtual-photon rate. In the searching of dark photons in HIC, we restrict to $m_U>2m_e$ so that the observable
signal corresponds to the on-shell two-body decay $U\to e^+e^-$.
The dilepton yield from dark photons then follows by multiplying the hadronic
$U$ yield by the visible branching fraction, while the partonic contribution is
directly rescaled:
\begin{align}
N_U &= \sum_h N_{h\to UX} + N_{q\bar q\to U},\\[3pt]
N^{U\to e^+e^-} &=
{\rm Br}(U\to e^+e^-)\sum_h N_{h\to UX}
\;+\;
\varepsilon^2\,N_{q\bar q\to e^+e^-}^{(M=m_U)},
\label{yieldtotal}
\end{align}
where $h$ represents the sum over all hadronic sources (\ref{decay1}-\ref{decay9}).
In Eq.~(\ref{yieldtotal}) we treat hadronic and partonic sources differently.
For hadronic channels, PHSD produces on-shell dark photons $U$, and the
observable dilepton yield is obtained by multiplying the produced $U$ yield by
the visible branching fraction ${\rm Br}(U\to e^+e^-)$.
For the partonic contribution, we approximate on-shell $U$ emission by mapping
the PHSD virtual-photon yield from $q\bar q\to\gamma^\ast\to e^+e^-$ at invariant
mass $M=m_U$ onto $q\bar q\to U\to e^+e^-$ in the narrow-width treatment; hence
the final-state $e^+e^-$ is already implicit, and the conversion amounts to a
simple $\varepsilon^2$ rescaling of the corresponding $\gamma^\ast$ yield.

For the partonic channel, we implement $U$ contribution by rescaling
the corresponding virtual-photon dilepton yield at $M=m_U$ by $\varepsilon^2$,
which already yields the $e^+e^-$ final state in the narrow-width treatment.
A detailed derivation of the partial-width ratios, kinematic factors, and their
numerical implementation in PHSD is provided in Ref.~\cite{Jorge:2025gph}.

Rare channels such as $K^+\to e^+\nu U$, $\Sigma^+\to p\,U$, or
$D^{*0}\to D^0U$, as well as production from primary Drell-Yan processes and
electromagnetic bremsstrahlung, are not included, since their rates are either
strongly constrained or subleading in the mass and energy ranges considered
here (see, e.g., Refs.~\cite{Pospelov:2008zw,Fabbrichesi:2020wbt,Ilten:2015hya}).

With the dark photon production channels and the total dilepton yield
Eq.~(\ref{yieldtotal}), PHSD provides the dark photon contribution to
the dilepton spectrum for any $(m_U,\varepsilon)$.  
We now use this prediction to convert the additional yield into quantitative
upper limits on the kinetic mixing parameter $\varepsilon^{2}(M_U)$.

\subsection{Extracting theoretical constraints on $\varepsilon^{2}(m_{U})$}
\label{sec:constraints-eps}

The kinetic mixing parameter $\varepsilon^{2}$ and the dark photon mass $m_{U}$
are a priori free parameters.  

Since PHSD provides the full invariant-mass spectrum from all SM
sources, any additional contribution from dark photon production must appear only as a small perturbation of the SM yield. Here, we only consider the decay of dark photons to dielectrons ($U \to e^+e^-$).

For each invariant-mass bin of width $\mathrm{d}M$ (here $\mathrm{d}M=10$~MeV),
we compute the integrated dark photon dielectron yield with $m_U$ in
$[m_U,m_U+\mathrm{d}M]$ and divide by the bin width, defining
\begin{equation}
\frac{\mathrm{d}N^{\mathrm{sum}U}}{\mathrm{d}M}
=
\sum_i \frac{N^{U\to e^+e^-}_i}{\mathrm{d}M},
\end{equation}
where the sum runs over all kinematically accessible production channels listed
in Eqs.~(\ref{decay1})-(\ref{decay10}). Because each hadronic source scales as
$\varepsilon^{2}$ and the partonic contribution is rescaled in the same way,
the total yield is linear in $\varepsilon^{2}$ \cite{Jorge:2025gph},
\begin{equation}
\label{eq:dNdm-eps-scaling}
\frac{\mathrm{d}N^{\mathrm{sum}U}}{\mathrm{d}M}
  \;=\;
\varepsilon^{2}\,
\frac{\mathrm{d}N^{\mathrm{sum}U}_{\varepsilon=1}}{\mathrm{d}M},
\end{equation}
where $\mathrm{d}N^{\mathrm{sum}U}_{\varepsilon=1}/\mathrm{d}M$ is obtained by
setting $\varepsilon=1$ in the production amplitudes, while keeping the
branching fractions (and hence the visible suppression in the invisible regime)
consistent with the chosen dark sector parameters.

\begin{figure*}[th]
    \centering    \includegraphics[width=0.99\linewidth]{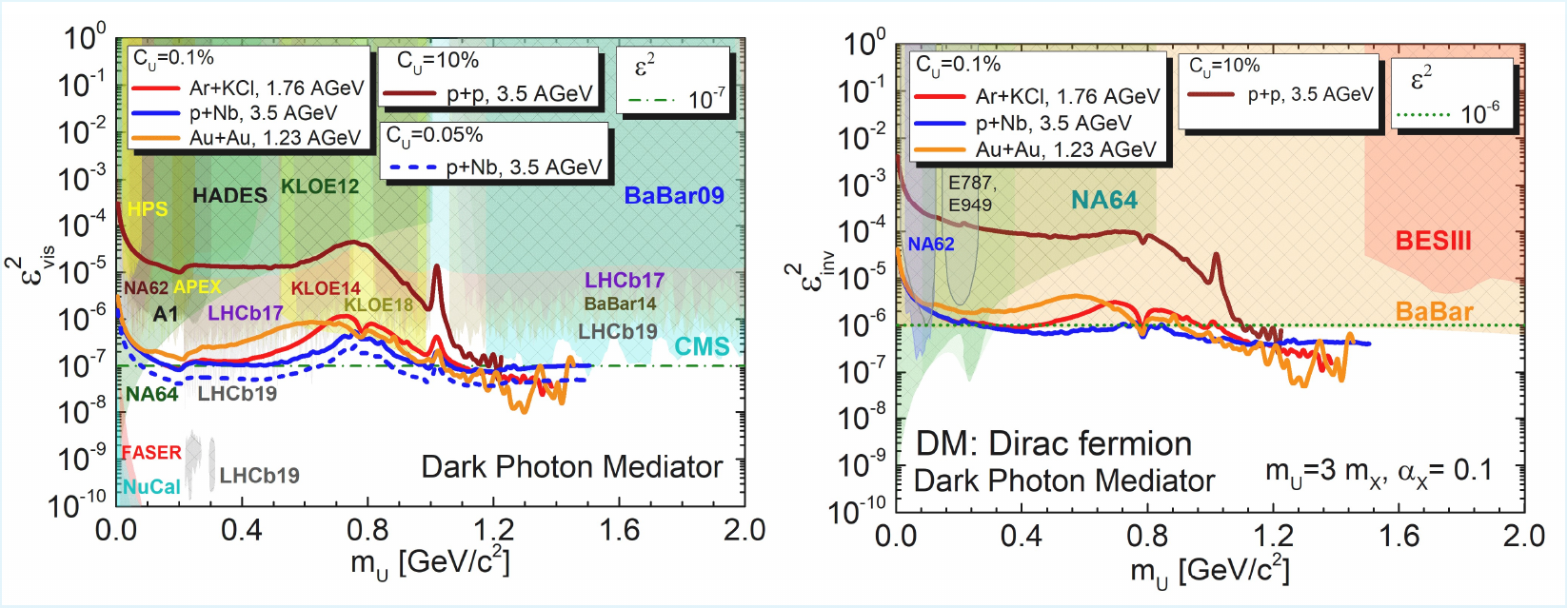}
\caption{
Upper limits on the kinetic mixing parameter for visible and invisible dark photon decays,
$\varepsilon^{2}_{\rm vis}(m_U)$ (left) and $\varepsilon^{2}_{\rm inv}(m_U)$ (right), 
extracted from the PHSD dilepton spectra in 
$p+p$ at $E_{\rm lab}=$  3.5~AGeV (brown), $Au+Au$ at $E_{\rm lab}=$ 1.23~AGeV (orange), 
$p+Nb$ at $E_{\rm lab}=$ 3.5~AGeV (blue), and $Ar+KCl$ at $E_{\rm lab}=$ 1.76~AGeV (red). In both panels, the PHSD kinetic mixing is constructed exclusively from the
$U\to e^+e^-$ decay mode, following Eq. \ref{eq:eps_bound_CU}.
We assume a Dirac fermion as the dark matter candidate for the invisible kinetic mixing.
For comparison, the left panel includes existing exclusion limits from 
HADES, HPS, NA62, NA64, APEX, A1, KLOE, BaBar, LHCb, FASER, NuCal and CMS \cite{Fabbrichesi:2020wbt,Ilten:2018crw}, 
while the right panel shows the leading invisible search bounds from 
NA64, E787/E949, BaBar and BESIII   \cite{BaBar:2017tiz,BESIII:2022oww,Crivelli:2023pxa}.
Horizontal green lines indicate benchmark values 
$\varepsilon^{2}=10^{-7}$ (dot dashed) in the left panel, 
and $\varepsilon^{2}=10^{-6}$ (dotted) in the right panel.
PHSD curves correspond to surplus parameters $C_U$ as indicated in the legends:
$C_U=0.1\%$ for all systems, with additional curves for 
$C_U=10\%$ in $p+p$ at $E_{\rm lab}=$ 3.5~AGeV (both panels) and 
$C_U=0.05\%$ in $p+Nb$ at $E_{\rm lab}=$ 3.5~AGeV (left panel). A Dirac fermion has been used as a DM candidate. 
}
\label{epsilon_phsd}
\end{figure*}

To quantify the maximal dark photon contribution still compatible with the
PHSD standard model (SM) dilepton yield in a given mass bin, we introduce a
surplus factor $C_U$. The parameter $C_U$ provides a measure of the excess over the PHSD SM yield, effectively capturing the
combined experimental precision (see Refs. \cite{Schmidt:2021hhs,Jorge:2025gph} for more details). We require, in each mass bin,

\begin{equation}
\frac{\mathrm{d}N^{\mathrm{sum}U}}{\mathrm{d}M}
\;\le\;
C_U\,
\frac{\mathrm{d}N^{\mathrm{sumSM}}}{\mathrm{d}M},
\label{eq:CU-def}
\end{equation}
where $\mathrm{d}N^{\mathrm{sumSM}}/\mathrm{d}M$ is the PHSD prediction from SM
sources alone.  Combining Eqs.~(\ref{eq:dNdm-eps-scaling}) and (\ref{eq:CU-def})
yields the bin-wise upper limit
\begin{equation}
\varepsilon^{2}
=
C_{U}
\left(
  \dfrac{\mathrm{d}N^{\rm sumSM}}{\mathrm{d}M}
  \;\middle/\;
  \dfrac{\mathrm{d}N^{\rm sumU}_{\varepsilon=1}}{\mathrm{d}M}
\right),
\label{eq:eps_bound_CU}
\end{equation}
which we evaluate across the mediator-mass range of interest.

For presentation, it is convenient to distinguish the two limiting regimes:
\begin{equation}
\varepsilon^{2} =
\begin{cases}
\varepsilon^{2}_{\rm vis}(m_U), & m_U < 2m_{\chi}, \\[6pt]
\varepsilon^{2}_{\rm inv}(m_U,m_\chi,\alpha_\chi), & m_U > 2m_{\chi},
\end{cases}
\label{eq:eps_piecewise_final}
\end{equation}
where $\varepsilon^{2}_{\rm vis}(m_U)$ corresponds to the peak search regime
with ${\rm Br}(U\to e^+e^-)\simeq 1$, while
$\varepsilon^{2}_{\rm inv}(m_U,m_\chi,\alpha_\chi)$ accounts for the suppression of the dilepton
signal through ${\rm Br}(U\to e^+e^-)\ll 1$ once $U\to\chi\bar\chi$ is open.
Equation~(\ref{eq:eps_bound_CU}) makes explicit that the mapping between a bound
on the dilepton excess and a bound on $\varepsilon^2$ depends on the dark sector
parameters that control $\Gamma_{\rm inv}$ (e.g.\ $\alpha_\chi$ and
$m_U/m_\chi$), as illustrated below.\\

Figure~\ref{epsilon_phsd} summarizes the limits on the kinetic mixing parameter 
obtained from PHSD dilepton spectra in the visible and invisible regimes using Eq. \ref{eq:eps_bound_CU}.  
The left panel displays the visible bounds, $\varepsilon^{2}_{\rm vis}(m_U)$, obtained under the assumption that the dark photon decays predominantly into standard model final states, in particular $U\!\to e^{+}e^{-}$. 
In this regime, a dark photon would appear as a narrow peak above the  SM 
dilepton spectra, so the extracted limits correspond to the maximal value of 
$\varepsilon^2$ for which the PHSD prediction remains consistent with the measured 
spectra.  
Exclusion bounds from fixed-target, collider, and beam-dump experiments   
have been added for comparison and shown as the hatched filled regions \cite{Fabbrichesi:2020wbt,Ilten:2018crw}.
A detailed discussion of the visible decay analysis and comparison with the experimental upper limits can be found in our previous work~\cite{Jorge:2025gph}.

The right panel displays the invisible bounds, $\varepsilon^{2}_{\rm inv}(m_U)$, 
obtained when the decay channel $U\!\to\chi\bar{\chi}$ is open and dominates the 
total width.  
The PHSD limits remain competitive across the sub-GeV mass range and are shown 
together with leading invisible search constraints from NA64, E787/E949, BaBar, and BESIII \cite{BaBar:2017tiz,BESIII:2022oww,Crivelli:2023pxa}.

Overall, the two panels highlight that dilepton measurements in heavy-ion and 
proton-nucleus collisions provide complementary sensitivity to visible and 
invisible dark photon scenarios.  
By covering mediator masses below and above hadronic resonances and by probing 
distinct decay regimes, the PHSD-based limits offer a unique contribution to the 
global dark photon parameter space.

 \begin{figure}
    \centering    \includegraphics[width=0.99\linewidth]{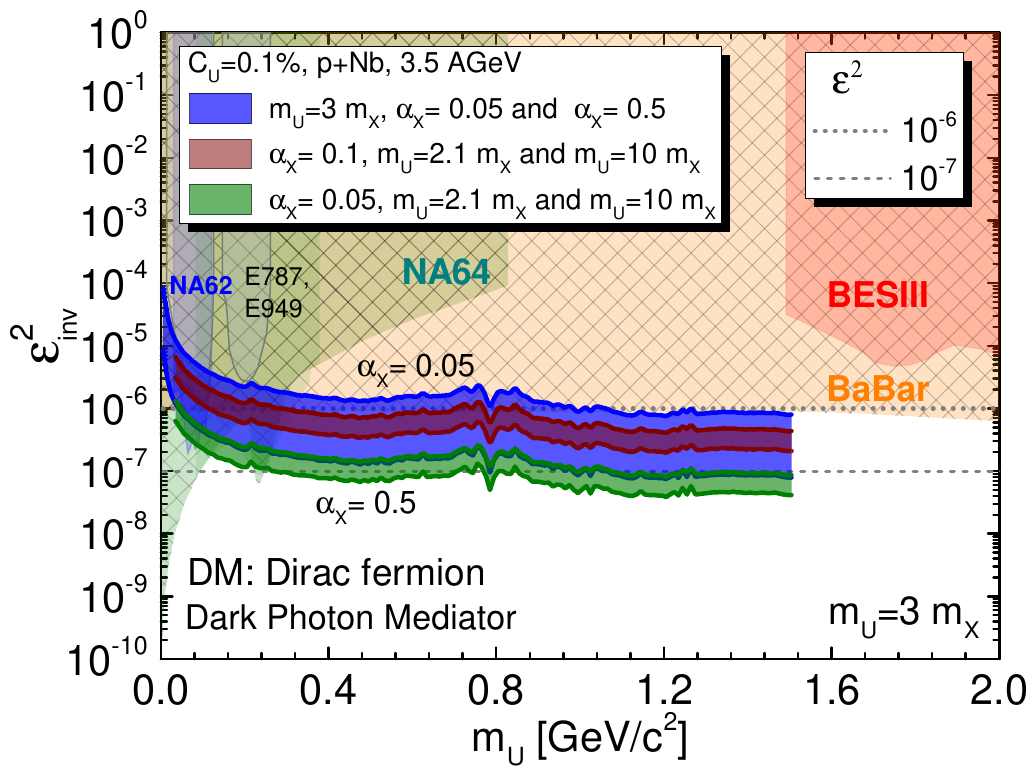}
\caption{Invisible decay constraints on the kinetic mixing parameter,
    $\varepsilon_{\rm inv}^{2}(m_U,m_\chi,\alpha_\chi)$, obtained from PHSD for 
    $p+Nb$ collisions at $E_{\rm lab}=$ 3.5~AGeV with surplus parameter $C_{U}=0.1\%$, assuming a Dirac fermion as dark matter candidate.
    The vertical axis shows the upper limit on $\varepsilon_{\rm inv}^{2}$ as a
    function of the mediator mass $m_{U}$, while the coloured bands illustrate
    the impact of varying the dark coupling $\alpha_{\chi}$ and the mass ratio
    $m_{U}/m_{\chi}$.
    The blue band corresponds to $m_{U}=3m_{\chi}$ with 
    $\alpha_{\chi}$ varied between $0.05$ and $0.5$.
    The red band shows $\alpha_{\chi}=0.1$ for two representative mass ratios,
    $m_{U}/m_{\chi}=2.1$ and $10$, and the green band displays the analogous
    variation for $\alpha_{\chi}=0.05$.
    Existing invisible search bounds from NA62, NA64, E787/E949, BaBar, and
    BESIII \cite{BaBar:2017tiz,BESIII:2022oww,Crivelli:2023pxa} are shown as hatched filled regions.
    Horizontal grey lines indicate reference values 
    $\varepsilon^{2}=10^{-6}$ (dotted) and $10^{-7}$ (dot-dashed).
    }
    \label{fig:epsilon_inv_scan}
\end{figure}

Figure~\ref{fig:epsilon_inv_scan} extends the invisible constraints of
Fig.~\ref{epsilon_phsd} (right) by explicitly scanning the dark sector
parameters $\alpha_{\chi}$ and $m_{U}/m_{\chi}$.  
For a fixed PHSD surplus $C_{U}=0.1\%$, each coloured band encodes how the
mapping from the visible PHSD yield to the underlying
$\varepsilon_{\rm inv}^{2}(m_U,m_\chi,\alpha_\chi)$ changes when the invisible width
$\Gamma_{\rm inv}(U\to\chi\bar{\chi})$ is modified.

The blue band illustrates the effect of varying the coupling
$0.05\le\alpha_{\chi}\le0.5$ at fixed ratio $m_{U}=3m_{\chi}$: increasing
$\alpha_{\chi}$ enhances $\Gamma_{\rm inv}$, reduces the visible branching
fraction $Br_{\rm vis}$, and therefore requires a larger $\varepsilon^{2}$ to
produce the same dilepton excess, pushing the upper edge of the band upward.
The red and green bands, instead, keep $\alpha_{\chi}$ fixed and bracket the
range between $m_{U}/m_{\chi}=2.1$ and $10$.  
For given $\alpha_{\chi}$, lowering $m_{\chi}$ (larger $m_{U}/m_{\chi}$)
again strengthens the invisible channel and weakens the PHSD limit, while
heavier dark matter leads to a smaller invisible width and correspondingly
stronger bounds (lower parts of the bands).

The tightest limits are obtained for the benchmark with small coupling and
heavy dark matter (green band, $\alpha_{\chi}=0.05$, $m_{U}=10m_{\chi}$),
which approaches the level $\varepsilon^{2}\sim10^{-7}$ over a broad mass
range.  
Because the PHSD yield scales linearly with the surplus parameter $C_{U}$, this
curve can equivalently be interpreted as corresponding to an effective excess
of $C_{U}\simeq0.01\%$ over the SM baseline, rather than $0.1\%$ as adopted in
Fig.~\ref{epsilon_phsd}.  
This comparison illustrates both the sensitivity of heavy-ion dilepton data to
invisible dark photon decays and the role played by the dark sector
parameters in translating the PHSD bounds into constraints on the fundamental
kinetic mixing.

In summary, the PHSD analysis provides upper limits on the kinetic mixing
parameter $\varepsilon^{2}$ in both the visible and invisible regimes,
for a range of dark sector benchmarks $(m_{\chi},m_{U},\alpha_{\chi})$.  
These bounds, however, do not by themselves guarantee a viable dark matter
scenario: the same mediator that controls the dark photon phenomenology also
governs the self-scattering of dark matter in astrophysical halos.  
In the following section, we therefore turn to Self-Interacting Dark Matter
(SIDM) and compute the velocity-dependent self-interaction cross section, in order to confront our benchmark models with
astrophysical constraints from dwarf galaxies, galaxies, and clusters.

\section{Self-Interacting Dark Matter}\label{sec:sidm}

The standard $\Lambda$ Cold Dark Matter ($\Lambda$CDM) paradigm, in which dark
matter (DM) is cold and effectively collisionless, successfully accounts for
large-scale cosmological observations, including the cosmic microwave background,
galaxy clustering, and gravitational lensing \cite{Frenk:2012ph}. 
On galactic and sub-galactic scales, however, several long-standing tensions
between collisionless $N$-body predictions and inferred inner halo density
profiles have been discussed. A prominent example is the core-cusp
problem: while high-resolution simulations typically yield centrally steep
profiles, $\rho(r)\propto r^{-1}$ \cite{Navarro:1996gj}, dwarf and
low-surface-brightness galaxies often favor approximately constant-density cores
in their inner regions \cite{deBlok:2009sp}.

Self-interacting dark matter (SIDM) provides a minimal and well-motivated
extension of $\Lambda$CDM in which DM particles undergo elastic scattering with
one another \cite{Tulin:2013teo}. For cross sections per unit mass in the
approximate range $\sigma/m_\chi \sim 0.1$- $10~\mathrm{cm^2\,g^{-1}}$,
self-interactions enable efficient heat transport and partial thermalization of
halo centers, thereby softening cusps into cores while leaving the large-scale
successes of $\Lambda$CDM essentially unchanged.

A central requirement is that the interaction strength depends on the typical
velocity scale of the system. Dwarf galaxies probe relative velocities of order
$\langle v\rangle\sim 50~\mathrm{km\,s^{-1}}$, where sizeable
self-interactions can produce kiloparsec-scale cores, whereas Milky-Way- sized
halos probe an intermediate regime,
$\langle v\rangle\sim 250~\mathrm{km\,s^{-1}}$. At larger mass scales,
galaxy groups and clusters reach substantially higher velocities,
$\langle v\rangle\sim 1150$- $1900~\mathrm{km\,s^{-1}}$, and therefore
impose stringent upper limits, since excessive scattering would noticeably
alter their shapes and internal dynamics. Visable SIDM scenarios must therefore
feature a pronounced suppression of $\sigma/m_\chi$ at group and cluster
velocities while allowing comparatively large values at dwarf scales
\cite{Colquhoun:2020adl}.

This scale dependence arises naturally if the self-interaction is mediated by a light force carrier. Here, we adopt a Yukawa potential, which provides a minimal description of the scattering process within the Born approximation and has been widely used in both analytical studies and numerical simulations \cite{Ibe:2009mk,Correa:2022dey,Correa:2023rzg}. This potential leads to a velocity-dependent transport cross section, $\sigma(v)$, that decreases at high velocities \cite{Feng:2009mn}. Other interaction models satisfying similar perturbative conditions are expected to exhibit qualitatively similar behavior. In the remainder of this section we adopt this framework for a dark photon mediator and compute the relevant velocity-dependent self-interaction cross sections, enabling a direct comparison with astrophysical constraints from dwarfs, galaxies, groups, and clusters.


\subsection{Calculations of self-interaction cross sections}
\label{subsec:calc_xs}

In this work, we compute the velocity-dependent transport cross sections with
the public code \textsc{CLASSICS} (\emph{CalcuLAtionS of Self-Interaction Cross
Sections})~\cite{Colquhoun:2020adl}, which provides fast and accurate solutions
of the non-relativistic scattering problem for Yukawa interactions,
\cite{Feng:2009hw,Tulin:2017ara},
\begin{equation}
  V(r)=\pm\,\frac{\alpha_\chi}{r}\,e^{-m_U r},
\end{equation}
where 
the upper (lower) sign corresponds to repulsive (attractive) interactions
between like charges.

For structure formation, the relevant transport quantities are the
momentum-transfer and viscosity cross sections
\cite{Khrapak:2014xqa,Tulin:2013teo,Tulin:2017ara},
\begin{align}
  \sigma_T(v) &= \int d\Omega\,(1-\cos\theta)\,\frac{d\sigma}{d\Omega},\\
  \sigma_V(v) &= \int d\Omega\,\sin^2\theta\,\frac{d\sigma}{d\Omega},
\end{align}
which suppress forward scattering and quantify the efficiency of momentum/heat
transport in halos.  
Here $v$ denotes the relative velocity
of the two scattering particles in
the center-of-mass frame,
$v \equiv |\mathbf{v}_1-\mathbf{v}_2| ,$
which is the appropriate non-relativistic variable controlling Yukawa
scattering.

The code can also account for the exchange symmetry of identical particles by
decomposing the scattering into even and odd partial waves. This symmetry decomposition is implemented for the
viscosity cross section $\sigma_V$, whose $\sin^2\theta$ weighting
emphasizes large-angle scattering and reduces sensitivity to the forward region,
facilitating a robust even/odd separation~\cite{Colquhoun:2020adl}.
Concretely, one defines $\sigma_{V}^{\rm even}(v)$ and $\sigma_{V}^{\rm odd}(v)$
from the partial-wave expansion of the underlying two-body scattering amplitude,
restricting to even or odd orbital angular momenta, respectively.
For unpolarized dark matter, the physical viscosity cross-section is obtained by
averaging over these symmetry channels, this leads to the
combinations
\begin{align}
  \sigma_{V}^{\rm scalar}(v) &= \sigma_{V}^{\rm even}(v),\\[4pt]
  \sigma_{V}^{\rm fermion}(v) &= \frac{3}{4}\,\sigma_{V}^{\rm odd}(v)
                           + \frac{1}{4}\,\sigma_{V}^{\rm even}(v),\\[4pt]
  \sigma_{V}^{\rm vector}(v)  &= \frac{1}{3}\,\sigma_{V}^{\rm odd}(v)
                           + \frac{2}{3}\,\sigma_{V}^{\rm even}(v).
\end{align}
where the weights reflect the multiplicities of antisymmetric and symmetric spin
configurations for $s=1/2$ and $s=1$ particles.

\subsection{Galaxy and group/cluster constraints} 
\label{subsec:groups_clusters} 

Galaxy groups and clusters provide an important laboratory for testing SIDM in the high-velocity regime, $\langle v\rangle \sim 10^3~\mathrm{km\,s^{-1}}$, where self-interactions must be sufficiently suppressed so as not to overly modify the inner density structure and halo shapes of massive systems. In this context, the literature reports both (i) values inferred from core-size measurements in relaxed halos within a given dynamical modelling, and (ii) complementary constraints based on the requirement that self-interactions remain compatible with observed halo properties.

In this work, we adopt the Jeans+lensing analysis of relaxed systems presented in Ref.~\cite{Sagunski:2020spe} as a phenomenological high-velocity criterion at $\langle v\rangle\sim 10^3~\mathrm{km\,s^{-1}}$. That analysis infers the interaction strength required to reproduce the measured core sizes of groups and clusters. Expressed in terms of a momentum-transfer cross section per unit mass, galaxy groups are consistent with $\sigma/m_\chi \sim \mathcal{O}(0.1\text{-}1)\, \mathrm{cm^2\,g^{-1}}$ at $\langle v\rangle \sim 10^3~\mathrm{km\,s^{-1}}$, while clusters favour somewhat smaller values. Operationally, we treat these inferred values as a conservative high-velocity viability band: parameter points predicting significantly larger self-interactions at group/cluster velocities are discarded, while points within (or below) the inferred band are retained for the combined analysis.

\begin{figure*}[t]
    \centering
    \includegraphics[width=0.49\linewidth]{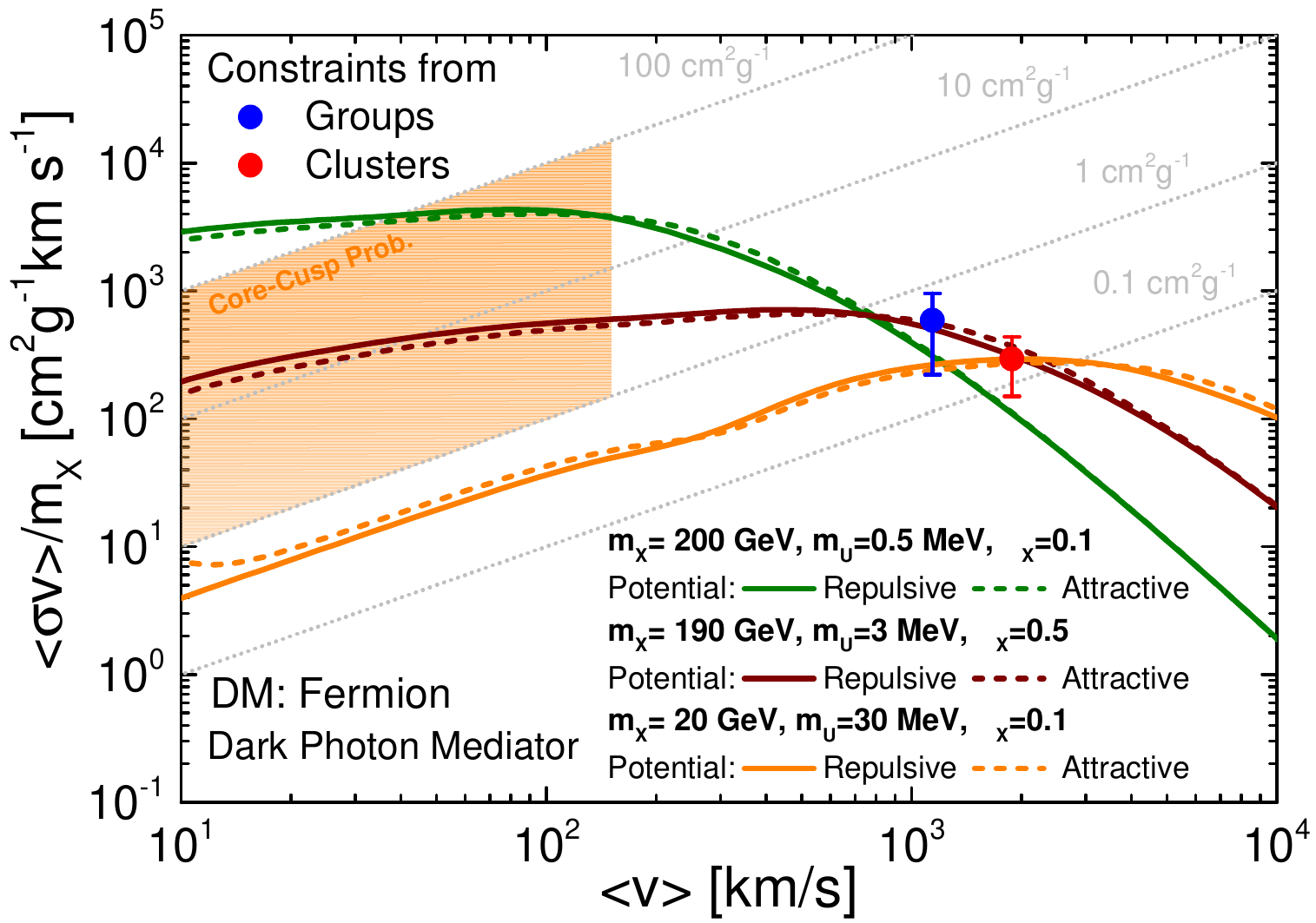}
    \includegraphics[width=0.49\linewidth]{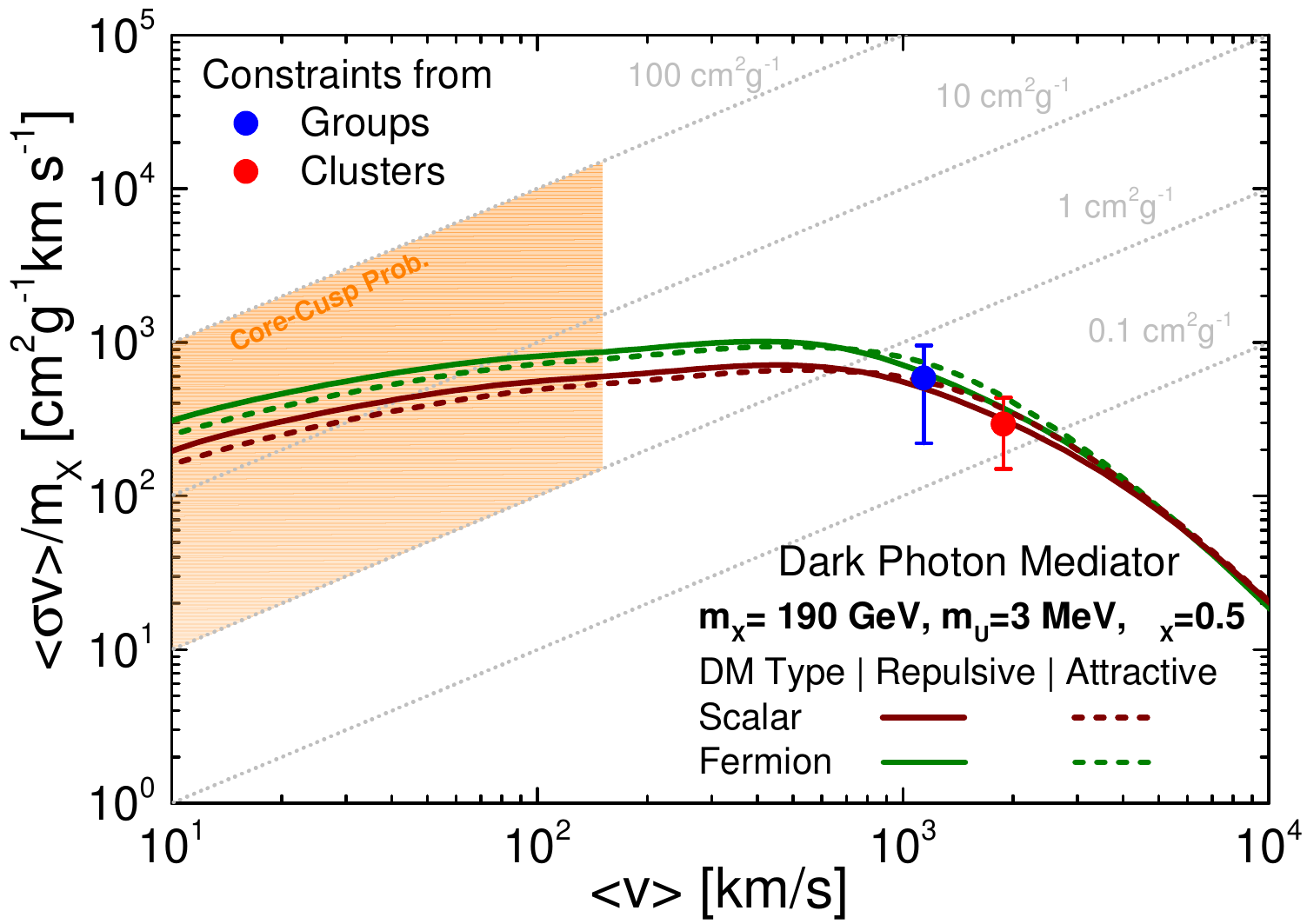}
  \caption{
    Velocity-averaged momentum-transfer cross section per unit mass, $\langle \sigma v \rangle/m_\chi$, as a function of the mean relative velocity
    $\langle v\rangle$, for Yukawa-mediated SIDM with a dark-photon mediator.  We added the group/cluster constraints taken from  Ref.~\cite{Sagunski:2020spe} for 
     direct comparison.
    Grey diagonal dotted lines correspond to constant values of
    $\sigma/m_\chi$ multiplied by $\langle v \rangle$.
    Left: Dirac-fermion dark matter for three benchmark choices of
    $(m_\chi,m_U,\alpha_\chi)$ shown in the legend; solid (dashed) lines denote
    repulsive (attractive) interactions.
    Right: comparison between scalar and fermionic dark matter for the benchmark
    shown in the legend, again distinguishing repulsive (solid) and attractive
    (dashed) potentials.
    Blue and red points with error bars indicate the inferred interaction
    strengths for galaxy groups and clusters, respectively, as compiled in
    Ref.~\cite{Sagunski:2020spe}; in our combined analysis, these points are used
    as a high-velocity viability band, effectively excluding significantly
    larger cross sections at group/cluster velocities.
    Grey diagonal lines correspond to constant values of
    $\sigma/m_\chi = 0.1,\,1,\,10$ and $100~\mathrm{cm^2\,g^{-1}}$.
    The orange band marks the approximate range
    $0.1 \lesssim \sigma/m_\chi \lesssim 10~\mathrm{cm^2\,g^{-1}}$ at
    $\langle v\rangle\sim 10$-$200~\mathrm{km\,s^{-1}}$ that is typically
    required to alleviate the core-cusp problem in dwarf and
    low-surface-brightness galaxies.
    All theory curves are computed with \textsc{CLASSICS}~\cite{Colquhoun:2020adl}.
}
    \label{fig:group_cluster_const}
    \vspace{0.3cm}
       \centering
    \includegraphics[width=0.49\linewidth]{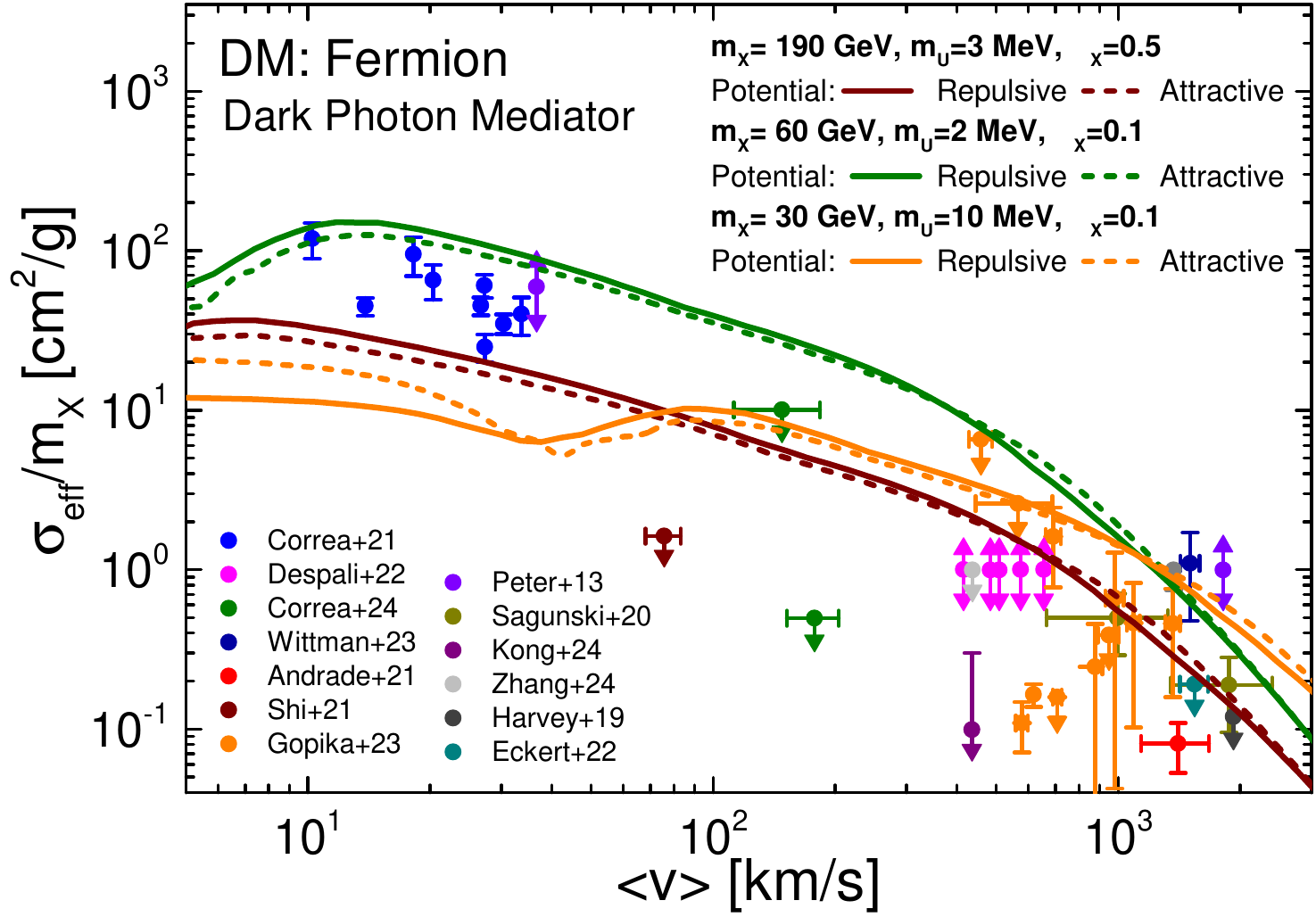}
    \includegraphics[width=0.49\linewidth]{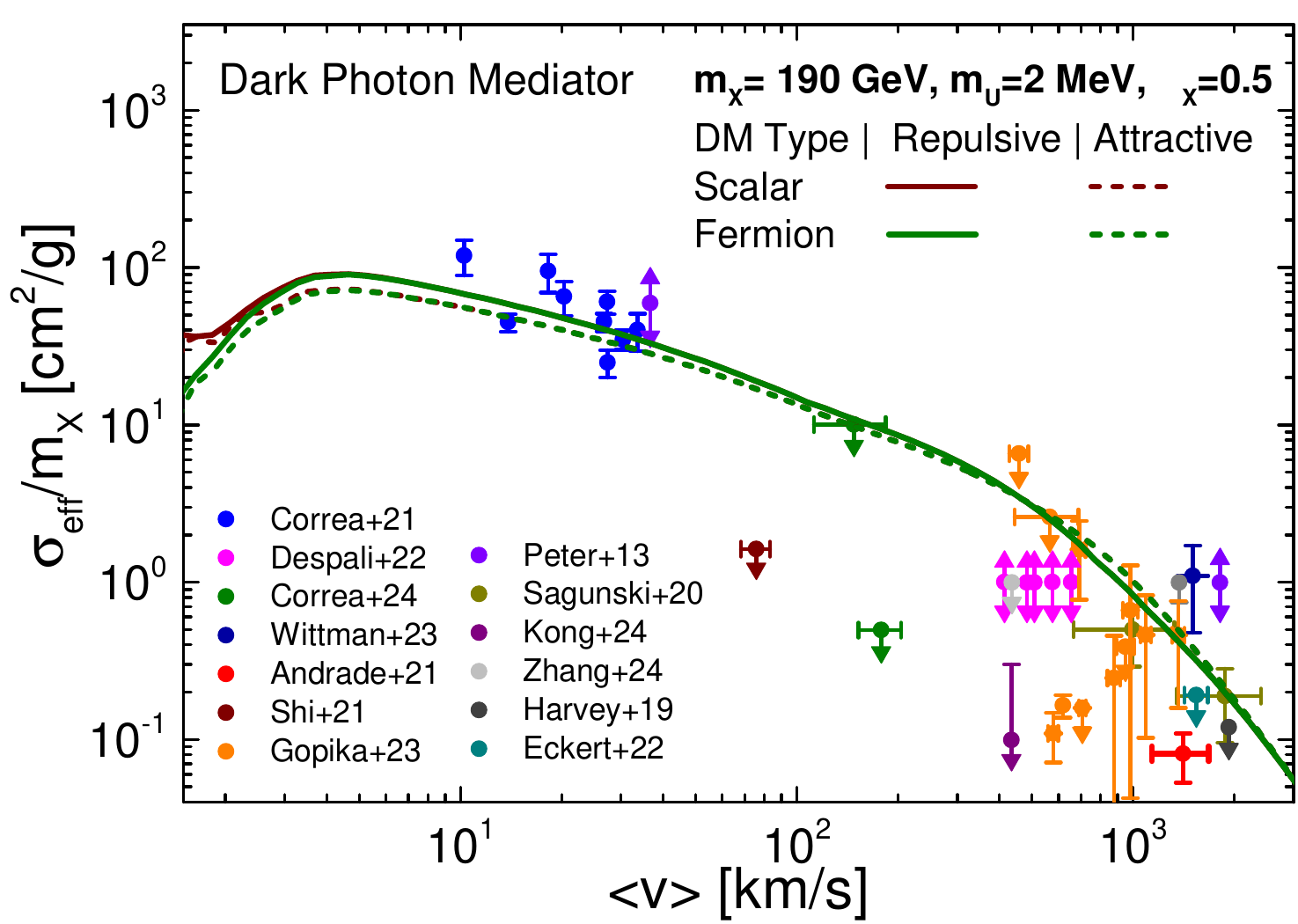}
    \caption{
    Effective self-interaction cross section per unit mass,
    $\sigma_{\mathrm{eff}}/m_\chi$, as a function of the mean relative velocity
    $\langle v\rangle$.
    The effective cross section $\sigma_{\rm eff}$ is constructed from the
    transfer-momentum cross section $\sigma$ and the relative velocity distribution as
    defined in Eq.~(\ref{eq:sigma_eff_text}), making it a more direct proxy for
    halo heat transport than $\sigma$.
    Solid lines show theoretical predictions obtained with \textsc{CLASSICS}
 for Yukawa-mediated SIDM with a dark photon
    mediator.
    Left: Dirac-fermion dark matter for several benchmark choices of
    $(m_\chi,m_U,\alpha_\chi)$ indicated in the legend, for both repulsive
    (solid) and attractive (dashed) potentials.
    Right: comparison between scalar and fermionic dark matter for the fixed
    benchmark $(m_\chi,m_U,\alpha_\chi)=(190~\mathrm{GeV},2~\mathrm{MeV},0.5)$,
    again distinguishing repulsive (solid) and attractive (dashed) interactions.
    Colored points represent astrophysical estimates and bounds on
    $\sigma_{\mathrm{eff}}/m_\chi$ from dwarf galaxies, Milky-Way-size halos,
    galaxy groups and clusters, compiled from Ref.~\cite{Fischer:2023lvl} and
    the \href{https://www.darkium.org}{\texttt{Darkium}} database.
    }
    \label{fig:eff_CS_const}
\end{figure*}

Fig.~\ref{fig:group_cluster_const} shows the velocity-weighted transport quantity
$\langle\sigma v\rangle/m_\chi$ as a function of the characteristic mean relative velocity
$\langle v\rangle$ for Yukawa-mediated SIDM with a dark-photon mediator.
The theory curves are obtained with the \textsc{CLASSICS} code~\cite{Colquhoun:2020adl},
which computes the velocity-dependent momentum-transfer cross section
$\sigma(v)\equiv\sigma_T(v)$ for Yukawa scattering.
To facilitate a direct comparison with the group/cluster inferences of
Ref.~\cite{Sagunski:2020spe}, which are quoted in terms of a Maxwell-Boltzmann averaged
transport cross-section, we use the quantity
$\langle\sigma v\rangle/m_\chi
\equiv
\bar{\sigma}/m_\chi\,\langle v\rangle,$
where $\bar{\sigma}_T$ is defined following Appendix~C of Ref.~\cite{Colquhoun:2020adl} as
\begin{equation}
\bar{\sigma}_T \equiv
\frac{\langle \sigma_T(v)\, v^2\rangle}
     {16\sqrt{2}\,v_0^2/\sqrt{\pi}},
\label{eq:sigmabarT_def_used}
\end{equation}
with angular brackets denoting an average over the Maxwell-Boltzmann relative velocity distribution.
The corresponding mean relative velocity is $\langle v\rangle = 4v_0/\sqrt{\pi}$.

For direct comparison, the blue and red points with error bars show the interaction strengths
inferred for galaxy groups and clusters, respectively, from the Jeans+lensing analysis of
Ref.~\cite{Sagunski:2020spe}.
The grey diagonal dotted lines indicate constant values of $\sigma/m_\chi$, obtained by dividing $\langle\sigma v\rangle/m_\chi$ by $\langle v\rangle$. Finally, the orange band indicates the range $0.1 \lesssim \sigma/m_\chi \lesssim 10~\mathrm{cm^2\,g^{-1}}$ at dwarf-galaxy velocities that is typically required to alleviate the core-cusp problem.

In the left panel of Fig.~\ref{fig:group_cluster_const} we consider Dirac-fermion dark matter and show
\textsc{CLASSICS} predictions for three benchmark choices of
$(m_U,m_\chi,\alpha_\chi)$ \cite{Colquhoun:2020adl}.  For each point, repulsive
(solid) and attractive (dashed) potentials yield very similar behaviour at
group/cluster velocities, where the scattering is already close to the Born-like
regime.  The curves pass through the group/cluster band near
$\langle v\rangle\sim 10^3~\mathrm{km\,s^{-1}}$ while remaining at or below
$\sigma/m_\chi\sim\mathcal{O}(1)\,\mathrm{cm^2\,g^{-1}}$, demonstrating
consistency with high-velocity halo constraints.  Extrapolating the same curves
to lower velocities shows that the interaction strength naturally increases
towards the dwarf-galaxy band, indicating that the same parameters can produce
efficient self-interactions in small halos while remaining viable at cluster
scales.

The right panel focuses on a single benchmark,
$(m_\chi,m_U,\alpha_\chi)=(190~\mathrm{GeV},3~\mathrm{MeV},0.5)$, and compares
scalar and fermionic dark matter.  For both repulsive and attractive
interactions, the scalar and fermion curves are nearly indistinguishable over
the velocity range probed by groups and clusters, differing only at the
$\mathcal{O}(1)$ level due to the distinct partial-wave decompositions.
This illustrates that, for mediator masses and couplings favoured by our
combined analysis, the high-velocity SIDM phenomenology is only weakly
spin-dependent, while still allowing sizable self-interactions at dwarf scales.

\subsection{Effective self-interaction cross section}
\label{subsec:eff_sigma}

Astrophysical systems over a wide range of masses and characteristic velocities
constrain the strength of dark matter self-interactions.  For a given halo
population, these constraints are commonly expressed in terms of an effective
transport cross-section per unit mass, evaluated at a characteristic relative
velocity $\langle v\rangle$ (see Sec.~\ref{subsec:calc_xs}).  Dwarf and
low-surface-brightness galaxies typically prefer
$\sigma_{\mathrm{eff}}/m_\chi \sim 0.1$-$10~\mathrm{cm^2\,g^{-1}}$ at
$\langle v\rangle\sim 20$-$80~\mathrm{km\,s^{-1}}$ in order to form
kiloparsec-scale cores, whereas galaxy groups and clusters predominantly provide
upper limits, requiring
$\sigma_{\mathrm{eff}}/m_\chi \lesssim \mathcal{O}(1)~\mathrm{cm^2\,g^{-1}}$ at
$\langle v\rangle\sim 10^3~\mathrm{km\,s^{-1}}$ so as not to overly round their
central density profiles.  In Fig.~\ref{fig:eff_CS_const} we represent these
measurements as colored points in the
$(\langle v\rangle,\sigma_{\mathrm{eff}}/m_\chi)$ plane using the compilation of
Ref.~\cite{Fischer:2023lvl} together with the \href{https://www.darkium.org}{\texttt{Darkium}} database.
To connect with these constraints, we therefore use the definition of an effective cross section~\cite{Fischer:2023lvl},
\begin{equation}
  \sigma_{\mathrm{eff}}
  \;=\;
  \frac{3}{2}\,
  \frac{\big\langle \sigma(v)\,v^{5}\big\rangle}{\big\langle v^{5}\big\rangle},
  \label{eq:sigma_eff_text}
\end{equation}
where angular brackets denote an average over the relative velocity distribution of the halo, which we approximate by a Maxwell-Boltzmann form.
The $v^{5}$ weighting follows from kinetic-theory considerations in the long-mean-free-path regime and highlights the range of relative velocities that contributes most efficiently to heat transport.
Here we construct $\sigma_{\rm eff}$ from the momentum-transfer cross section, $\sigma_T(v)$.
When an even/odd partial-wave decomposition is needed for the spin averages discussed above, then we use the viscosity cross section $\sigma_V(v)$.

Figure~\ref{fig:eff_CS_const} shows $\sigma_{\mathrm{eff}}/m_\chi$ as a function
of $\langle v\rangle$ for Yukawa-mediated SIDM with a dark photon mediator.
In the left panel, we consider Dirac fermion dark matter and display
\textsc{CLASSICS} predictions for several benchmark
choices of mediator mass $m_U$, dark matter mass $m_\chi$, and coupling
$\alpha_\chi$ (see legend), for both repulsive (solid) and attractive (dashed)
potentials.  The velocity dependence reflects the transition from the
non-perturbative regime at low velocities to the Born-like regime at high
velocities: the interaction strength can reach
$\sigma_{\mathrm{eff}}/m_\chi\sim 0.1$-$10~\mathrm{cm^2\,g^{-1}}$ at dwarf
velocities, while being suppressed at group/cluster velocities and remaining
compatible with the corresponding upper limits.

The right panel focuses on the benchmark
$(m_\chi,m_U,\alpha_\chi)=(190~\mathrm{GeV},2~\mathrm{MeV},0.5)$ and compares
scalar and fermionic dark matter.  Both spins exhibit a very similar velocity
dependence, with modest normalization differences arising from the distinct
partial-wave structure.

Overall, these results demonstrate that Yukawa-mediated SIDM models with
$\alpha_\chi=\mathcal{O}(0.1$-$0.5)$ can simultaneously yield sizeable
self-interactions at dwarf-galaxy velocities, as typically preferred by core
data, while remaining consistent with the predominantly upper-limit constraints
from galaxy groups and clusters.

\section{Thermal relic abundance}
\label{subsec:thermal_relic}

The present-day abundance of a cold thermal relic is set by chemical freeze-out
in the early Universe. For a dark matter species $\chi$ with mass $m_\chi$ and
thermally averaged effective (co)annihilation rate $\langle\sigma_{\rm eff}v\rangle$,
it is convenient to work with the comoving number density
$Y \equiv n_\chi/s$, where $n_\chi$ is the physical number density of $\chi$ and $s$ is
the entropy density. Introducing the dimensionless inverse temperature
$x \equiv m_\chi/T$, with $T$ the temperature of the thermal bath, the Boltzmann
equation reads~\cite{Kolb:1990vq,Gondolo:1990dk,Pospelov:2007mp},
\begin{equation}
\frac{dY}{dx}
= -\,\frac{s\,\langle\sigma_{\rm eff} v\rangle}{H\,x}\,(Y^2-Y_{\rm eq}^2),\label{BE_comoving}
\end{equation}
where $Y_{\rm eq}(x)\equiv n_{\chi,{\rm eq}}(T)/s(T)$ is the equilibrium yield.
The entropy density is
$s=(2\pi^2/45)\,g_{*s}(T)\,T^3$, with $g_{*s}(T)$ the effective number of
relativistic degrees of freedom in entropy, and during radiation domination the
Hubble rate is
$H=\sqrt{4\pi^3 g_*(T)/45}\,T^2/M_{\rm Pl}$, where $g_*(T)$ is the corresponding
energy-density degrees of freedom and $M_{\rm Pl}$ denotes the (non-reduced)
Planck mass.
In the non-relativistic regime ($x\gg 1$), $Y_{\rm eq}\propto x^{3/2}e^{-x}$, and
freeze-out occurs once the annihilation rate per particle,
$\Gamma_{\rm ann}=n_\chi\langle\sigma_{\rm eff}v\rangle$, drops below the Hubble
expansion rate.

This formalism applies both to elastic and inelastic dark sectors.  In general,
$\langle\sigma_{\rm eff}v\rangle$ denotes the thermally averaged effective
rate controlling chemical decoupling, including the appropriate combination of
annihilation and, when relevant, coannihilation channels, as well as any
velocity dependence from thresholds and resonant effects in the underlying
matrix elements.  

Requiring the solution of the Boltzmann equation to reproduce the observed relic
abundance, ~\cite{Planck:2018vyg,WMAP:2012nax}
\begin{equation}
 \Omega_\chi h^2\simeq 0.12  \pm 0.001  ,
\end{equation}
 selects a narrow
band in parameter space.  This band is often referred to as the thermal
target.  For conventional weakly coupled scenarios it corresponds to the
well-known benchmark scale
$\langle\sigma_{\rm eff}v\rangle\sim\mathcal{O}(10^{-26})~\mathrm{cm^3\,s^{-1}}$,
up to $\mathcal{O}(1)$ variations due to the particle-physics realization and
the temperature dependence of $\langle\sigma_{\rm eff}v\rangle$.  

In the following, we determine the thermal target  for the dark photon 
framework \cite{Alexander:2016aln,Battaglieri:2017aum,Krnjaic:2025noj} considered in this work and confront them with laboratory constraints
on the kinetic mixing, as well as with the astrophysical requirements implied
by self-interacting dark matter.

\subsection{Relic density calculation}
\label{subsec:reddeliver}

To compute the thermal relic abundance in vector-mediated dark-sector scenarios,
we employ the public \textsc{ReD-DeLiVeR} code~\cite{Foguel:2024lca}, which provides
a self-consistent evaluation of the thermally averaged annihilation rate
$\langle\sigma v\rangle(T)$ including mediator-width effects and the relevant
Standard-Model thresholds.
Mediator partial widths and branching ratios are implemented in a data-driven
manner, ensuring that threshold openings, the transition from leptonic to
hadronic channels, and resonance effects are consistently reflected in the
thermal average.

In this work, we focus on elastic dark matter scenarios, corresponding to a
single stable species coupled to the dark photon.
We consider three representative realizations of the dark sector:
(i) a Dirac fermion,
(ii) a Majorana fermion, and
(iii) a complex scalar.
For the fermionic cases, the Majorana and Dirac limits are treated separately,
while in all cases, the relic abundance is governed by standard $2\to2$
annihilation processes.

The dominant freeze-out channel is annihilation into Standard-Model fermions via an
$s$-channel dark photon,
$\chi\bar{\chi}\to f\bar f$ (or $\varphi^\ast\varphi\to f\bar f$ for scalar DM).
In the non-relativistic limit, Dirac fermions typically admit an $s$-wave contribution,
whereas complex scalars and Majorana DM are $p$-wave suppressed, $\sigma v \propto v^2$.
This spin-dependent velocity scaling is automatically captured by the matrix elements
implemented in \textsc{ReD-DeLiVeR}.

The thermal evolution is described by a single Boltzmann equation for the
comoving abundance (Eq. \ref{BE_comoving}), with an effective annihilation cross section
$\langle\sigma v\rangle$ that fully incorporates mediator width effects and the
velocity dependence induced by resonant annihilation near $m_U\simeq2m_\chi$ 

For each point in the parameter space
$(m_\chi,m_U,\varepsilon,g_{\chi})$,
\textsc{ReD-DeLiVeR} solves the freeze-out dynamics and computes the relic
density $\Omega_\chi h^2$.
The corresponding thermal target curves, defined as the set of parameter points
that reproduce the observed relic abundance, $\Omega_{\rm DM}h^2\simeq0.12$, are used
throughout this work as a reference to compare with constraints from heavy-ion
collisions, accelerator and beam-dump experiments, as well as astrophysical
limits from self-interacting dark matter.

In selecting thermal-target points, we require the computed relic abundance to
lie within a finite acceptance band around the Planck value. Unless stated
otherwise, we adopt$
\big|\Omega_\chi h^2 - 0.12\big| < \Delta_{\rm relic}$ ,
with $\Delta_{\rm relic}$ chosen according to the numerical scan resolution in
\textsc{ReD-DeLiVeR}. In our runs we use $\Delta_{\rm relic}$ in the range
$10^{-3}$--$10^{-2}$ (and up to $\mathcal{O}(10^{-2})$ in narrow regions where
hadronic thresholds induce rapid variations), and we have verified that the
qualitative conclusions are stable under tightening this criterion in smooth
regions.

Unless stated otherwise, each thermal relic curve shown below is obtained by
fixing the dark coupling $\alpha_\chi$ and a mass hierarchy
$R\equiv m_U/m_\chi$, and then scanning the mediator mass $m_U$.
Along such a curve, the dark matter mass therefore varies as $m_\chi=m_U/R$.
Accordingly, each point on a given $(m_U,\varepsilon)$ (or $(m_U,\varepsilon^2)$)
target corresponds to a distinct pair $(m_\chi,m_U)$ linked by the chosen
hierarchy $R$.

\begin{figure}[t]
    \centering
    \includegraphics[width=0.99\linewidth]{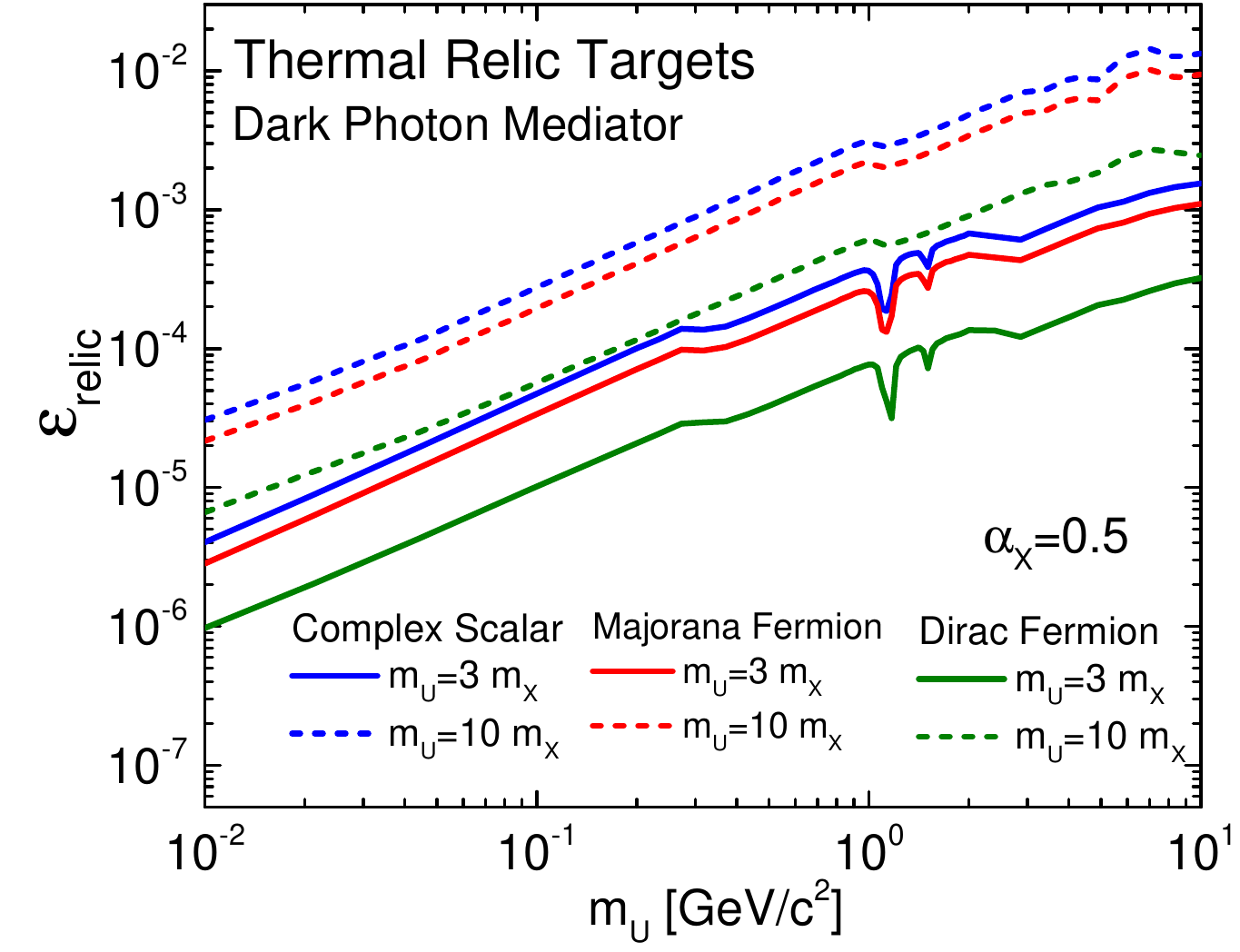}
    \caption{
Thermal relic target curves for the kinetic mixing parameter $\varepsilon$ as
a function of the mediator mass $m_U$ at fixed dark coupling $\alpha_\chi=0.5$,
obtained with \textsc{ReD-DeLiVeR}~\cite{Foguel:2024lca} by imposing the
relic density condition $\Omega_\chi h^2\simeq0.12$.
For each curve we fix the mass hierarchy $R\equiv m_U/m_\chi$, taking $R=3$
(solid) and $R=10$ (dashed), and scan $m_U$; the dark matter mass varies along
the curve as $m_\chi=m_U/R$.
Colours denote the DM spin assignment: complex scalar (blue), Majorana fermion
(red), and Dirac fermion (green).
Localized dips arise from hadronic thresholds and resonance structures in the
visible width entering the thermal average. 
   }
     \label{fig:thermal_targets_alpha05}
\end{figure}

\begin{figure*}[ht]
    \centering
    \includegraphics[width=0.49\linewidth]{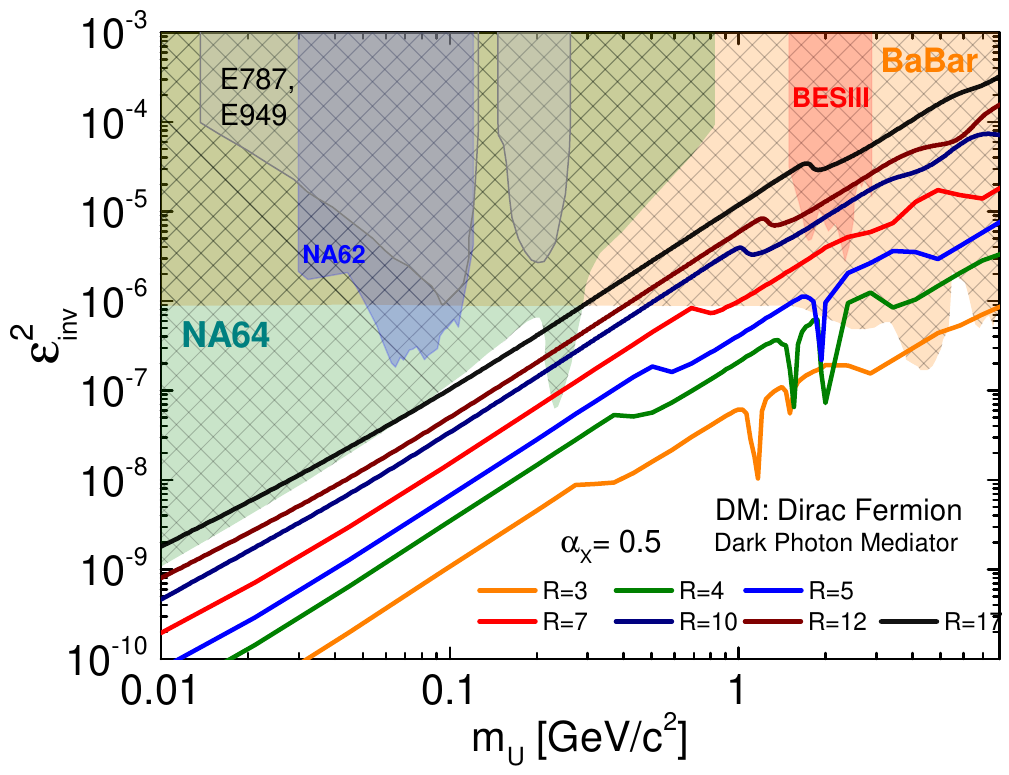}
    \includegraphics[width=0.49\linewidth]{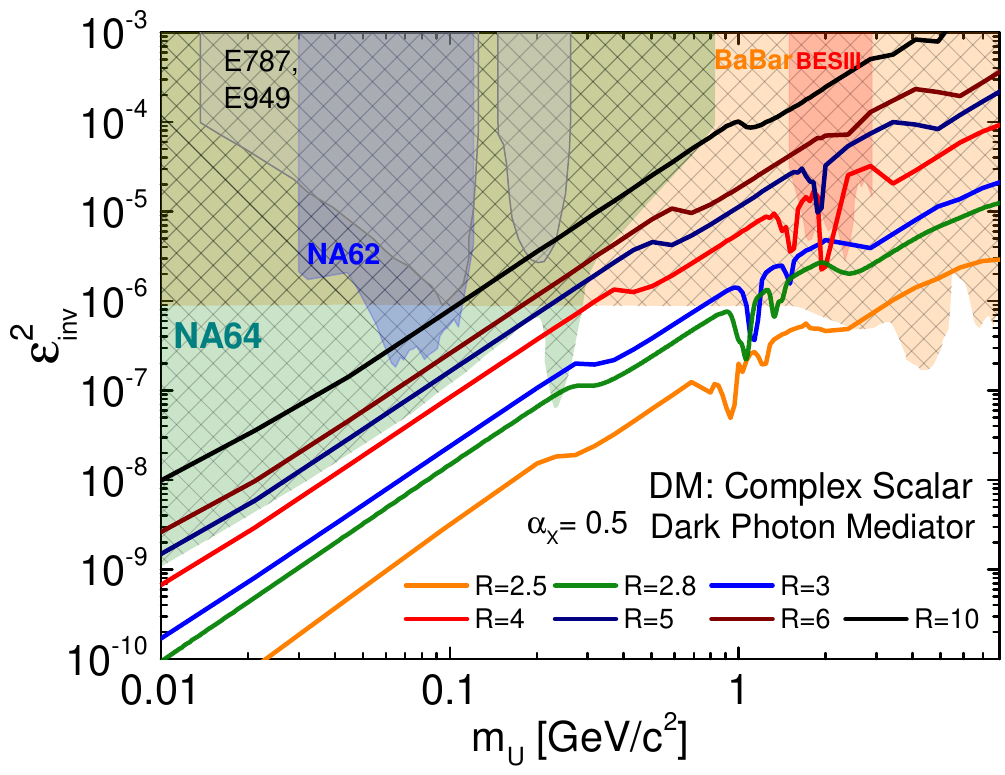}
\caption{
Invisible search limits on the kinetic mixing parameter $\varepsilon^{2}_{\rm inv}$
compared to thermal relic targets for a dark photon mediator.
Shaded regions show existing constraints from NA64, NA62,
E787/E949, BaBar, and BESIII in the $\varepsilon^{2}_{\rm inv}$-$m_U$ plane.
Solid curves are thermal relic targets computed with \textsc{ReD-DeLiVeR}
by imposing $\Omega_\chi h^2\simeq 0.12$.
In both panels we fix $\alpha_\chi=0.5$ and, for each coloured line, fix the
mass hierarchy $R\equiv m_U/m_\chi$ while scanning $m_U$; equivalently,
$m_\chi=m_U/R$ varies along the target.
Left: Dirac-fermion DM targets for $R=3,4,5,7,10,12,17$.
Right: complex-scalar DM targets for $R=2.5,2.8,3,4,5,6,10$.
Localized dips arise from hadronic thresholds in the mediator width.
}
  \label{fig:eps_inv_targets}
\end{figure*}

Figure~\ref{fig:thermal_targets_alpha05} shows the thermal relic target lines in
the $(m_U,\varepsilon)$ plane for a fixed dark coupling $\alpha_\chi=0.5$.
For each dark matter spin assignment, complex scalar (blue), Majorana fermion
(red), and Dirac fermion (green), we display two representative mass
hierarchies, $R=m_U/m_\chi = 3$ (solid) and $10$ (dashed).
Each curve corresponds to the value of the kinetic mixing $\varepsilon(m_U)$
required to reproduce the observed relic abundance $\Omega_\chi h^2\simeq0.12$,
as computed with \textsc{ReD-DeLiVeR}; for fixed $R$, scanning $m_U$ implies
$m_\chi=m_U/R$ along the target.

The overall rise of the curves with $m_U$ reflects the fact that, for a heavier
mediator, a larger mixing is needed to maintain the canonical annihilation rate.
The relative ordering encodes the spin dependence of the annihilation kernel:
for fixed $m_U$ and $R$, Dirac fermions typically require the smallest
$\varepsilon$, Majorana fermions an intermediate value, and complex scalars the
largest coupling to achieve the same relic density.
This pattern is expected since Dirac annihilation is typically dominated by an
$s$-wave contribution, while Majorana and complex scalar annihilation are often
$p$-wave suppressed; therefore, for the same mass scale, a $p$-wave scenario
generally needs a slightly larger $\varepsilon$ to reach $\Omega_\chi h^2\simeq0.12$
because the effective annihilation rate decreases roughly as $1/x$ as the Universe
cools ($x\equiv m_\chi/T$).
Narrow dips around $m_U\sim\mathcal{O}(1~\mathrm{GeV})$ arise from the combined
effect of (i) hadronic thresholds entering the mediator width and (ii) resonant
$s$-channel annihilation close to the kinematic condition $m_U\simeq 2m_\chi$.
These thermal targets provide the benchmark against which we compare the PHSD
limits on $\varepsilon$ obtained from heavy-ion dilepton spectra in the visible
and invisible regimes.

\begin{figure*}[ht]
    \centering
    \includegraphics[width=0.49\linewidth]{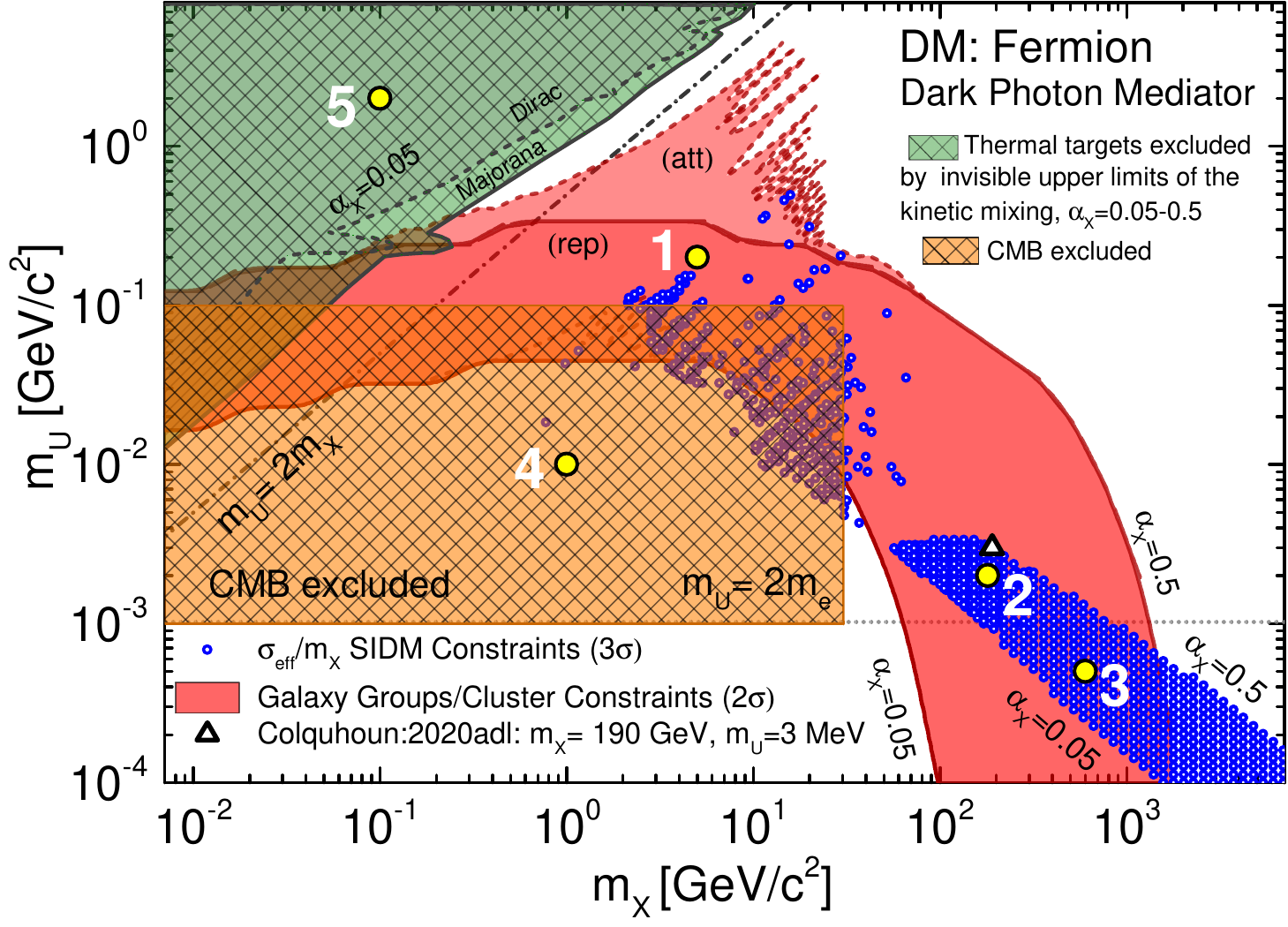}
    \includegraphics[width=0.49\linewidth]{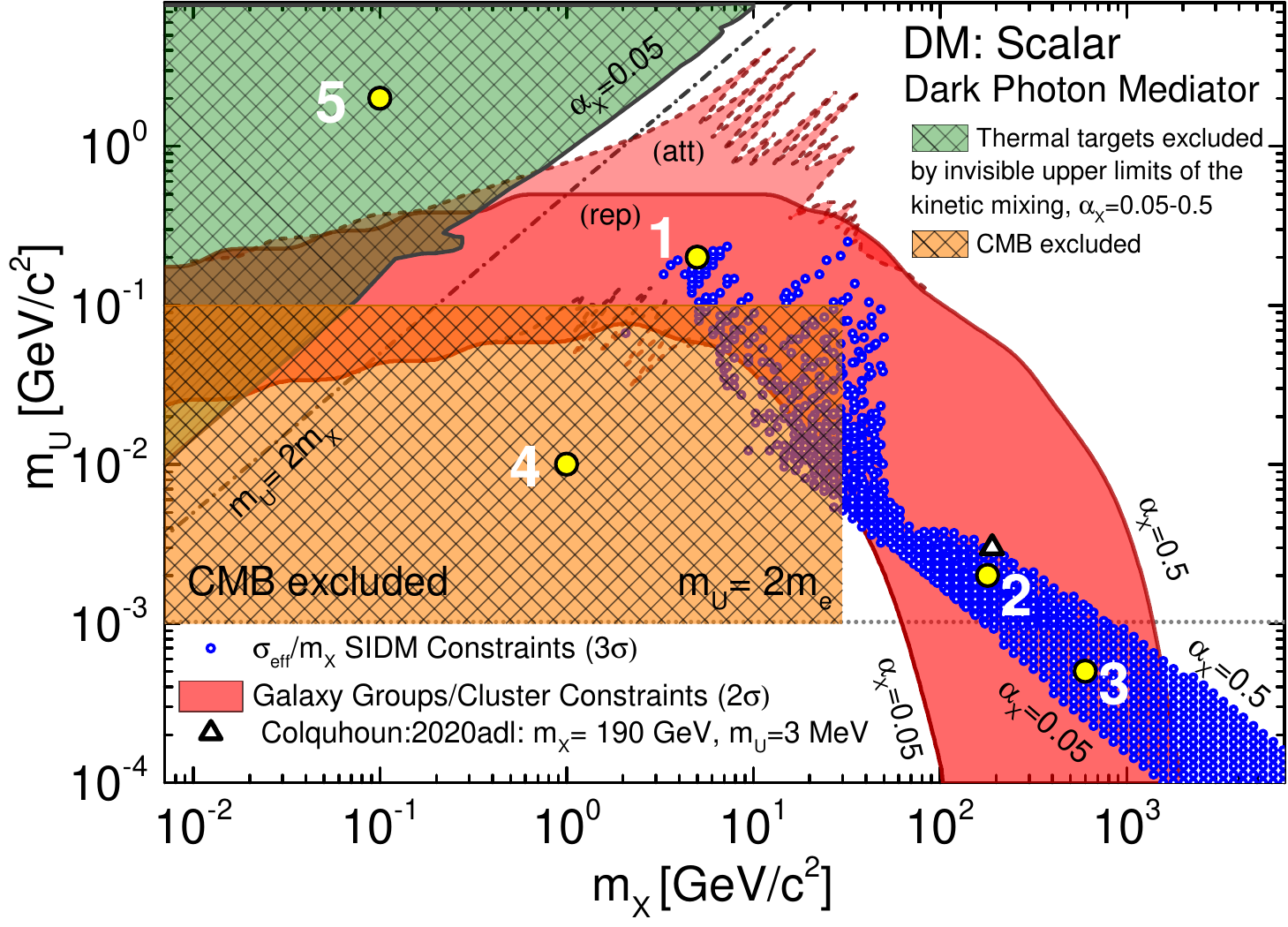}
\caption{
Combined constraints in the $(m_\chi,m_U)$ plane for dark matter coupled to a
dark photon mediator.
Left: fermionic DM (Dirac or Majorana); right: complex scalar DM.
Blue circles indicate parameter points where the Yukawa interaction reproduces the effective
self-interaction cross-section constraints $\sigma_{\rm eff}/m_\chi$ from dwarfs, galaxies,
groups, and clusters ($3\sigma$), while the red band shows the subset consistent with the
group/cluster constraints at $2\sigma$.
The orange hatched region at low masses is excluded by CMB data \cite{Kaplinghat:2013yxa, Planck:2018vyg}, where
dark matter radiation interactions would distort recombination and the CMB anisotropies.
Green hatched regions mark thermal relic target points (for $\alpha_\chi = 0.05$-$0.5$)
that are ruled out by the invisible upper limits on the kinetic mixing extracted from PHSD
dilepton spectra.
The diagonal line labelled $m_U = 2m_\chi$ denotes the kinematic threshold for
$U \to \chi\bar{\chi}$, and the horizontal line at $m_U = 2m_e$ marks the opening of the
$e^{+}e^{-}$ channel.
Dotted curves labelled $\alpha_\chi = 0.05$ and $0.5$ illustrate typical coupling values,
and the black triangle indicates the benchmark $m_\chi = 190~\mathrm{GeV}$,
$m_U = 3~\mathrm{MeV}$ of Ref.~\cite{Colquhoun:2020adl}.
Five benchmark points (yellow circles): BP1-BP5 are highlighted in  in the figure and defined in the text.
}
    \label{fig:mu_mx_planes}
\end{figure*}

Figure~\ref{fig:eps_inv_targets} shows how the thermal relic targets in the
invisible regime are constrained by existing bounds on the kinetic mixing.
Each coloured line corresponds to the value of $\varepsilon^{2}_{\rm relic}(m_U)$
required to reproduce the observed dark matter abundance for a fixed coupling
$\alpha_\chi=0.5$ and a fixed mass ratio $R=m_U/m_\chi$, for Dirac fermion DM
(left) and complex scalar DM (right).
Whenever a thermal relic curve enters one of the shaded exclusion regions from
NA64, NA62, E787/E949, BaBar, or BESIII~\cite{BaBar:2017tiz,BESIII:2022oww,Crivelli:2023pxa},
the corresponding point in parameter space is ruled out: the model would require
a kinetic mixing larger than experimentally allowed to achieve the correct relic
density.
Equivalently, these intersections map into excluded bands in the $(m_\chi,m_U)$
plane at fixed $\alpha_\chi$ for each DM spin assignment, and only the segments
of the thermal targets lying below all invisible limits are retained in our
subsequent scans.

\section{$m_U(m_\chi)$ parameter space}\label{sec:mu_mx_planes}

Figure~\ref{fig:mu_mx_planes} summarizes our combined constraints in the
$(m_\chi,m_U)$ plane for fermionic (left) and complex-scalar (right) dark matter
interacting through a kinetically mixed dark photon.
The different overlays encode, in a compact form, the interplay between
(i) halo-scale self-interactions, (ii) early-Universe relic density targets,
(iii) CMB limits at low masses, and (iv) heavy-ion bounds in the invisible regime.

The blue points indicate parameter choices for which the Yukawa-mediated
self-interactions satisfy the effective self-interaction requirements compiled
from dwarfs, Milky-Way-size halos, groups, and clusters, expressed in terms of
$\sigma_{\rm eff}/m_\chi$ as described in
Secs.~\ref{subsec:calc_xs}-\ref{subsec:eff_sigma}.
These points yield sizeable interactions at dwarf velocities (supporting core
formation) while remaining suppressed at group/cluster velocities, where the
constraints are predominantly upper limits. The scan includes both signs of the
Yukawa potential, i.e.\ attractive and repulsive interactions.

The red band highlights the subset of blue points that additionally satisfies
the group/cluster constraint at the $2\sigma$ level as implemented in
Sec.~\ref{subsec:groups_clusters}; operationally, it delineates the
high-velocity-viable portion of the SIDM parameter space used in our subsequent
combined scans. Within this region, repulsive and attractive potentials are
distinguished by dark and light red shading, respectively. As expected for
attractive Yukawa scattering in the resonant/non-perturbative regime, the
attractive case exhibits localized resonant enhancements, visible as narrow
structures around $m_\chi\sim 1$-$20~\mathrm{GeV}$ and
$m_U\sim 0.01$-$1~\mathrm{GeV}$.

The orange hatched region at low $(m_\chi,m_U)$ is excluded by CMB constraints
for $m_\chi \lesssim 30~\mathrm{GeV}$ and $1 \lesssim m_U \lesssim 100~\mathrm{MeV}$
(see Ref.~\cite{Kaplinghat:2013yxa} and also
Refs.~\cite{Lopez-Honorez:2013cua,Planck:2018vyg,Krnjaic:2025noj}).
In this region, symmetric dark matter can still annihilate during recombination,
$\chi\bar \chi \to U U$.
Since the mediator decays mainly into electrons
(${\rm Br}(U\to e^+e^-)\simeq 1$), it injects electromagnetic energy into the plasma.
This alters the ionization history and is incompatible with the observed CMB anisotropies.

To project the heavy-ion constraints into the $(m_\chi,m_U)$ plane, we evaluate
the PHSD upper limits on the kinetic mixing in the invisible regime,
$\varepsilon^2_{\rm inv}(m_U)$, and compare them to the kinetic mixing required
to reproduce the observed relic abundance for each scanned point,
$\varepsilon^2_{\rm relic}(m_U,m_\chi,\alpha_\chi)$.
The green hatched regions mark thermal relic parameter points for which
$\Omega_{\rm DM}h^2\simeq 0.12$ is obtained (computed with \textsc{ReD-DeLiVeR}
for $\alpha_\chi=0.05$-$0.5$), but which satisfy
$\varepsilon^2_{\rm relic}(m_U,m_\chi,\alpha_\chi)>\varepsilon^2_{\rm inv}(m_U,m_\chi,\alpha_\chi)$
and are therefore excluded by the PHSD invisible limits.
Equivalently, these regions indicate where a standard thermal history would
predict an invisible decay scenario requiring kinetic mixing above the PHSD limit.

The diagonal dot-dashed line $m_U=2m_\chi$ marks the kinematic threshold above which the
mediator can decay invisibly to $\chi\bar\chi$, while the horizontal dotted line
$m_U=2m_e$ denotes the opening of the $e^+e^-$ channel.
The black triangle indicates the benchmark point $m_\chi=190~\mathrm{GeV}$,
$m_U=3~\mathrm{MeV}$ from Ref.~\cite{Colquhoun:2020adl}.

To guide the discussion, we mark five benchmark points (BP1-BP5) in
Fig.~\ref{fig:mu_mx_planes}.
BP1-BP3 are chosen inside the combined-allowed region to represent the three
phenomenological regimes highlighted by our scan:
\begin{enumerate}
  \item \textbf{BP1: light mediator, intermediate DM.}
  A sub-GeV mediator with $m_U\simeq 0.2~\mathrm{GeV}$ and
  $m_\chi\simeq 5~\mathrm{GeV}$.
  This point lies below the invisible threshold ($m_U<2m_\chi$), so dilepton
  searches operate in the visible regime and production is dominated by hadronic
  sources (Dalitz decays, vector mesons, and baryon resonances), providing a clean
  target for precision low-mass spectra.

  \item \textbf{BP2: ultra-light mediator, heavy DM.}
  An MeV-scale mediator with $m_U\simeq 2~\mathrm{MeV}$ and
  $m_\chi\simeq 180~\mathrm{GeV}$, representative of a strongly hierarchical
  Yukawa-mediated SIDM realization.
  In this setup, the self-interaction exhibits pronounced velocity dependence:
  it can be sizable at dwarf and group velocities while remaining sufficiently
  suppressed at cluster scales.
  This benchmark is compatible with
  both the group/cluster constraints (as in Ref. \cite{Colquhoun:2020adl}) and the bounds based on the effective
  self-interaction cross section. 

  \item \textbf{BP3: long-lived mediator, heavy DM.}
  A benchmark with a very light mediator, $m_U<2m_e$, with
  $m_U\simeq 0.5~\mathrm{MeV}$ and $m_\chi\simeq 600~\mathrm{GeV}$, for which all
  visible decays into charged SM states are kinematically forbidden and the
  invisible channel $U\to\chi\bar\chi$ is closed.
  Consequently, the mediator is effectively long-lived on detector scales; its
  phenomenology is governed by production via kinetic mixing and by loop-induced
  radiative decays (e.g.\ $U\to 3\gamma$), which are typically extremely suppressed
  in this mass range \cite{McDermott:2017qcg}.
\end{enumerate}

In addition, BP4 and BP5 are placed in excluded regions to illustrate, within
the same $(m_\chi,m_U)$ map, which condition excludes them:
\begin{enumerate}
 \setcounter{enumi}{3}
  \item \textbf{BP4: CMB-excluded point.}
  A representative low-mass configuration in the  excluded domain
  (orange hatched), chosen at $m_\chi\simeq 1~\mathrm{GeV}$ and
  $m_U\simeq 10~\mathrm{MeV}$.
  In this region, energy injection from residual annihilation around
  recombination can significantly modify the ionization history and hence the
  CMB temperature and polarization anisotropies, rendering this parameter point
  incompatible with CMB constraints.

  \item \textbf{BP5: PHSD-invisible excluded thermal point.}
A representative thermal relic configuration in the green hatched region,
chosen at $m_\chi\simeq 0.1~\mathrm{GeV}$ and $m_U\simeq 2~\mathrm{GeV}$ with
$m_U>2m_\chi$, such that invisible decays are kinematically allowed.
For these masses, achieving $\Omega_{\rm DM}h^2\simeq 0.12$ within the scanned
range $\alpha_\chi = 0.05$-$0.5$ requires a kinetic mixing $\varepsilon^2$
that exceeds the PHSD-derived invisible upper limits, rendering this point
incompatible with the heavy-ion constraints.

\end{enumerate}

Taken together, the two panels show that the simultaneous requirements of (i)
efficient self-interactions at dwarf scales, (ii) suppression at group/cluster
velocities, (iii) consistency with CMB bounds, and (iv) compatibility with the
invisible heavy-ion limits substantially restrict the viable parameter space.
In our scan, the permitted region is typically concentrated at light mediators
in the MeV-sub-GeV range and comparatively heavy dark matter, from a few GeV up
to the TeV scale. In this regime, Yukawa screening and the transition toward
the Born (weak-coupling) limit suppress the interaction strength at high
velocities while allowing enhanced scattering in small halos. Differences between the fermionic and scalar panels are driven primarily by the
spin dependence of the annihilation kernel (and hence the thermal targets) and
by the corresponding invisible widths, while the overall structure of the
allowed region remains similar.

Altogether, our results indicate that the sub-threshold region $m_U<2m_\chi$ is
typically favored in the combined analysis, since the mediator cannot decay
invisibly and, for $m_U>2m_e$, remains directly visible in dileptons with an
unsuppressed branching fraction. For $m_U>2m_\chi$ invisible decays reduce
${\rm Br}(U\to e^+e^-)$ and tighten the viable thermal relic parameter space once
the invisible limits are applied, while the very light-mediator domain
$m_U<2m_e$ corresponds to a long-lived scenario that is not probed by the
dilepton channel.

\section{Summary}
\label{sec:summary}

We have investigated a vector portal dark sector in which a kinetically mixed
dark photon $U$ couples to the standard model electromagnetic current with
strength $\varepsilon e$ and to a stable, symmetric, and elastic dark sector particle through a dark
gauge coupling $g_\chi$ (equivalently $\alpha_\chi$). This defines a parameter
space $(m_U,\varepsilon,m_\chi,\alpha_\chi)$ that connects heavy-ion dilepton
observables to thermal freeze-out and to velocity dependent self-interactions
relevant for structure formation. We considered three benchmark realizations of
the dark sector state: a Dirac fermion, a Majorana fermion (axial vector
coupling), and a complex scalar.

On the high-energy collision side, we used the PHSD transport approach, including an
on-shell dark photon produced from hadronic Dalitz channels, baryon-resonance
transitions, direct vector-meson decays, kaon decays, and partonic
$q\bar q\to U$ annihilation, followed by $U\to e^+e^-$ decays. Taking the PHSD standard model dilepton spectrum as a reference, we derived upper limits on the kinetic mixing,
$\varepsilon^2(m_U)$ for  the invisible regime ($m_U>2m_\chi$), where
$U\to\chi\bar\chi$ suppresses the dilepton branching fraction.

Furthermore, we incorporated cosmological and astrophysical requirements in two complementary
ways. First, we computed Yukawa-mediated self-interaction cross sections using
\textsc{CLASSICS} and constructed the effective transport quantity
$\sigma_{\rm eff}/m_\chi$ used throughout our halo level analysis. Confronting
these predictions with compilations of constraints and inferences from dwarfs,
Milky-Way-size halos, groups, and clusters, we identified regions in which
self-interactions are sufficiently large at dwarf velocities to support core
formation while remaining suppressed at group/cluster velocities. Second, we obtained thermal relic target curves with \textsc{ReD-DeLiVeR} by imposing
$\Omega_{\rm DM}h^2\simeq0.12$, incorporating mediator width effects and hadronic thresholds.

Combining these ingredients, we mapped the viable parameter space in the
$(m_\chi,m_U)$ plane for each DM spin assignment (Fig.~\ref{fig:mu_mx_planes}).
In the invisible regime, we project the heavy-ion constraints by comparing the
kinetic mixing required by freeze-out, $\varepsilon^2_{\rm relic}(m_U,m_\chi,\alpha_\chi)$,
to the invisible upper limits $\varepsilon^2_{\rm inv}(m_U,m_\chi,\alpha_\chi)$, and we
find that sizable portions of the would be thermal relic parameter space are
already excluded. In addition, CMB bounds remove a broad low-mass region where energy injection around recombination would distort the measured anisotropies.
 The simultaneous requirement of SIDM-like scattering at
dwarf scales and suppression at cluster scales further narrows the allowed
region, typically favoring a light mediator in the MeV-sub-GeV range and heavier
dark matter from a few tens of GeV up to the TeV scale.

To facilitate the interpretation of the combined constraints, we introduced five
benchmark points (BP1-BP5) in Fig.~\ref{fig:mu_mx_planes}. BP1-BP3 are chosen
within the combined-allowed region and represent (i) a sub-GeV mediator with
intermediate-mass DM in the visible regime, (ii) an ultra-light MeV mediator with
heavy DM typical of hierarchical SIDM realizations, and (iii) a long-lived
mediator scenario with $m_U<2m_e$ and $m_U<2m_\chi$, where both visible and
invisible two-body decays are kinematically forbidden and only highly suppressed
loop-induced channels remain. BP4 and BP5 are placed in excluded regions to
illustrate, within the same plane, the impact of CMB exclusions at low masses and
the removal of thermal relic points in the invisible regime by the PHSD
$\varepsilon^2_{\rm inv}(m_U)$ limits.

These findings favor the region $m_U<2m_\chi$ with $m_U>2m_e$, where the mediator remains
visible in dileptons, while the very light-mediator domain $m_U<2m_e$ corresponds to a long-lived
scenario not directly probed by the dilepton channel. Taken together, precision dilepton
measurements in heavy-ion collisions provide a complementary and competitive probe of
vector-portal dark sectors when interpreted jointly with thermal freeze-out and SIDM
phenomenology. Future improvements in low-mass dilepton data, extended system/energy coverage,
and refined astrophysical inferences will further sharpen the combined constraints and help
prioritize the most promising regions for experimental searches.

\begin{acknowledgments}
A.R.J. expresses gratitude for the financial support from the Stiftung Giersch. We also acknowledge the support by the Deutsche Forschungsgemeinschaft (DFG) through the grant CRC-TR 211 "Strong-interaction matter under extreme conditions" (Project number 315477589 - TRR 211) and the CNRS Helmholtz Dark Matter Lab (DMLab). The computational resources utilized for this work were provided by the Center for Scientific Computing (CSC) at Goethe University Frankfurt. G.-W.Y. acknowledge support from the University of Trento and the Provincia Autonoma di Trento (PAT, Autonomous Province
of Trento) through the UniTrento Internal Call for Research 2023 grant “Searching for Dark Energy off the beaten track” (DARKTRACK, grant agreement no. E63C22000500003), and from the Istituto Nazionale di Fisica Nucleare (INFN) through the Commissione Scientifica Nazionale 4 (CSN4) Iniziativa Specifica “Quantum Fields in Gravity, Cosmology and Black Holes” (FLAG).
\end{acknowledgments}

\bibliography{references}

\end{document}